\documentclass[fleqn,usenatbib]{mnras}

\usepackage{newtxtext,newtxmath}
\usepackage{graphicx}
\usepackage{subfigure}
\usepackage{xcolor}

\usepackage{mathtext,bm,bbm,amsmath,amsfonts,amssymb,indentfirst,syntonly,graphicx,epsf}
\usepackage{mathtools}
\usepackage[english]{babel}
\usepackage{calc}
\usepackage{tikz}
\usepackage[T1]{fontenc}
\usepackage{ae,aecompl}
\usepackage[normalem]{ulem}
\newcommand{\hc}{$H_0$}
\newcommand{\hcunit}{km~s$^{-1}$~Mpc$^{-1}$}
\def\hst{{\it HST}}

\title[Improved lens modelling and $H_0$ with transients]{Improved time-delay lens modelling and $H_0$ inference with transient sources}
\author[Ding et al.]
{Xuheng Ding$^{1,2}$,
Kai Liao$^{3}$\thanks{E-mail: liaokai@whut.edu.cn},
Simon Birrer$^{4}$,
Anowar J. Shajib$^{5}$,
Tommaso Treu$^{2}$,\newauthor
and Lilan Yang$^{6}$\\\\\\\\
$^{1}$Kavli IPMU (WPI), UTIAS, The University of Tokyo, Kashiwa, Chiba 277-8583, Japan\\
$^{2}$Department of Physics and Astronomy, University of California, Los Angeles, CA, 90095-1547, USA\\
$^{3}$School of Science, Wuhan University of Technology, Wuhan 430070, China\\
$^{4}$Kavli Institute for Particle Astrophysics and Cosmology and Department of Physics, Stanford University, Stanford, CA 94305,
USA\\
$^{5}$Department  of  Astronomy  \&  Astrophysics,  University  of Chicago, Chicago, IL 606374 \\
$^{6}$School of Physics and Technology, Wuhan University, Wuhan 430072, China
}
\begin{document}

\date{Accepted xxxx; Received xxxx; in original form xxxx}

\pagerange{\pageref{firstpage}--\pageref{lastpage}} \pubyear{2021}

\maketitle

\label{firstpage}

\begin{abstract}
Strongly lensed explosive transients such as supernovae, gamma-ray bursts, fast radio bursts, and gravitational waves are very promising tools to determine the Hubble constant ($H_0$) in the near future in addition to strongly lensed quasars.
In this work, we show that the transient nature of the point source provides an 
advantage over quasars: the lensed host galaxy can be observed before or after the transient's appearance. Therefore, the lens model can be derived from images free of contamination from bright point sources. 
We quantify this advantage by comparing the precision of a lens model obtained from the same lenses with and without point sources. Based on
Hubble Space Telescope (\hst) Wide Field Camera 3 (WFC3)
observations with the same sets of lensing parameters, we simulate realistic mock datasets of 48 quasar lensing systems (i.e., adding AGN in the galaxy center) and 48 galaxy--galaxy lensing systems (assuming the transient source is not visible but the time delay and image positions have been or will be measured). We then model the images and compare the inferences of the lens model parameters and $H_0$. We find that the precision of the lens models (in terms of the deflector mass slope) 
is better by a factor of 4.1 for the sample without lensed point sources, resulting in an increase of \hc\ precision by a factor of 2.9. 
The opportunity to observe the lens systems without the transient point sources provides an additional advantage for time-delay cosmography over lensed quasars. It facilitates the determination of higher signal-to-noise stellar kinematics of the main deflector, and thus its mass density profile, which in turn plays a key role in breaking the mass-sheet degeneracy and constraining \hc. 
\end{abstract}


\section{Introduction}

The Hubble constant ($H_0$) is currently the object of great attention. 
Reported values show a significant tension between the early-Universe and late-Universe measurements. Assuming a $\Lambda$ cold dark matter ($\Lambda$CDM) cosmology, the most precise constraints are taken from the early-Universe measurement of cosmic microwave background (CMB) observations from {\it Planck}, which is $H_0 = 67.4 \pm 0.5$~\hcunit~\citep{Planck2020}.
By using the CMB temperature and polarization anisotropy from the Atacama Cosmology Telescope and combining with large-scale information from WMAP, \citet{ACT2020} measured $H_0 = 67.6 \pm 1.1$~\hcunit\ in excellent agreement with the {\it Planck} result.
The Supernovae, \hc, for the Equation of State of dark energy (SH0ES) team, which uses a late-Universe probe by calibrating the type Ia supernovae (SN) distance ladder using Cepheids and parallax distances, measures $H_0 = 73.0 \pm 1.4 $~\hcunit~\citep{Riess2021}; it is in $4.2\sigma$ tension with the measurement by {\it Planck}.
Alternatively, using the distance ladder method with SN Ia and the tip of the red giant branch (TRGB), the Carnegie-Chicago Hubble Program (CCHP) measures $H_0 = 69.6 \pm 1.9$~\hcunit~\citep{Freedman2019, Freedman2020}. However, the SH0ES team measures a value of $H_0 = 72.4 \pm 2.0$~\hcunit using the TRGB stars to calibrate the  distance ladder~\citep{Yuan2019}.
This so-called ``Hubble tension'' is of great importance. If unknown systematic errors in the measurements can be excluded, solving it would require new physics beyond $\Lambda$CDM~\citep[e.g.,][]{Knox2020}.
Given the high stakes, multiple independent techniques are a vital safeguard against unknown unknowns.

Strong gravitational lensing provides a one-step method to determine $H_0$, since the time delay measurements between multiple images encode the information of absolute distances~\citep{Refsdal1964, Treu2016, Suyu2017}, independent of all other methods. Typically, the lensing system used for time-delay cosmography consists of a distant quasar and a foreground galaxy. The source quasar is lensed into multiple images that arrive at the observer at different times. Since quasar luminosity is variable, the delay between the images can be measured and turned into cosmological distances.
Lensed quasars are the traditional targets because they are the most abundant variable lensed sources known to date. However, as survey capabilities improve, other transient lensed sources are being discovered \citep{Kelly2015} and will be discovered in increasingly large numbers~\citep{Oguri2010}.

To infer $H_0$ using strong lensing systems, one needs at least three ingredients: 
1) time delays between images measured using light curve pairs; 
2) Fermat potential differences between the lensed images determined by high-resolution imaging of the lensed arcs, and spectroscopic kinematics; 
3) the distribution of mass along the line of sight to the lensed source.
For lensed quasars, current measurements have precision at the several percent level per system~\citep{Wong2020, Millon2020, Birrer2020}.

Much progress has been achieved in time-delay cosmography by the H0LiCOW \citep{Suyu2017,Wong2020}, COSMOGRAIL \citep{Courbin2017}, STRIDES \citep{Shajib2020}, and SHARP \citep{Chen2019} collaborations \citep[now combined in the TDCOSMO collaboration][]{Millon2020}, reaching $\sim 2$\% precision on \hc\ under the assumption that the radial mass density profile of the deflector can be described by a power-law mass profile (yields $H_0 = 74.2 \pm 1.6 $~\hcunit) or a composite of \citet[][hereafter NFW]{Navarro1997} dark matter halo and stars as described by the surface brightness scaled by a constant stellar mass to light ratio (yields $H_0 = 74.0 \pm 1.7 $~\hcunit).
Relaxing those assumptions on the radial mass density profile, the study by \citet[][hereafter TDCOSMO-IV]{Birrer2020} uses stellar kinematics to break the mass-sheet degeneracy and finds $H_0 = 74.5 \substack{+5.6 \\ -6.1} $~\hcunit; the precision of \hc\ drops from 2\% to 8\% using the 7 lenses in TDCOSMO, without changing the mean inferred \hc\ significantly. TDCOSMO-IV then introduces a hierarchical framework and combines the TDCOSMO lenses with external datasets to enhance the precision. Under the assumption that the deflectors of TDCOSMO and lenses in the Sloan Lens ACS (SLACS) survey are drawn from the same population, TDCOSMO-IV combines TDCOSMO+SLACS and achieves $H_0 = 67.4 \substack{+4.1 \\ -3.2} $~\hcunit with 5\% precision. The inferred mean value is shifted and offset to the TDCOSMO-only value towards the direction of the {\it Planck} \hc, although it still statistically agrees with the previous TDCOSMO measurement of \hc\ given the uncertainties. This shift of the mean could be either real or due to the invalidation of the assumption that the deflectors of TDCOSMO and SLACS are similar.

The current error budget of the analysis without radial mass density profile assumptions is dominated by statistical uncertainties in the kinematics. Thus, increasing the number of time-delay lensing systems will shrink the uncertainty on $H_0$, until the current systematic floor is reached. Examples of potential systematic floor include the so-called time-delay microlensing effect \citep{Tie2018, Liao2020}, line of sight corrections, substructure, and selection effects~\citep{Collett2016, Gilman2020} all estimated to be at around the percent level. Efforts are currently underway to increase the size of samples useful for time-delay cosmography in order to reach the same $\sim$2\% level of precision without radial profile assumptions~\citep{Birrer2020a}. 

Given the progress on time-domain surveys, time-delay cosmography will soon be able to utilize transients as an additional and abundant class of time-variable multiply imaged sources. As described in a recent review by \citet{Oguri2019}, for example, supernovae of all types~\citep{Oguri2010, Petrushevska2016, Petrushevska2018, Petrushevska2018a, Suyu2020}, gamma-ray bursts/afterglows in all bands, repeated fast radio bursts~\citep{Li2018}, and even gravitational waves with electromagnetic counterparts~\citep{Liao2017}, are expected to be regularly observed by large facilities like the Rubin Observatory Legacy Survey of Space and Time (LSST) in the optical~\citep{Oguri2010, Goldstein2017}, the Square Kilometer Array (SKA) in the radio~\citep{Wagner2019} and third-generation gravitational wave (GW) detectors such as the Einstein Telescope~\citep{Biesiada2014,Piorkowska2013,Ding2015,LiSS2018,Yang2019}\footnote{\citet{Hannuksela2019} recently searched for lensed GW signals in the ``Gravitational-wave  Transient  Catalog  1''~\citep{GWTC1} following the approach proposed by~\citet{Haris2018}; however, nothing was found. The possibility of the pair of GW170104 and GW170814 as lensed event was also estimated by~\citet{Dai2020} using the framework based on~\citet{Bar10}, but with very weak evidence.}.

The first observation of a multiply imaged supernovae~\citep{Kelly2015} brought us new insights and stimulated numerous studies of the detection~\citep{Goldstein2017,Shu2018,Rydberg2020}, the nature of SNe, and time-delay cosmography~\citep{Goldstein2018,Huber2020,Pierel2020}. Supernova Refsdal should soon provide the first competitive measurement of \hc\ from a multiply imaged transient~\citep{Grillo2018,Grillo2020}. The observed lensed SNe were well analyzed~\citep{Dhawan2020,Mortsell2020}.

In addition to being a way to increase sample size, strongly lensed transients can provide significant advantages over lensed quasars in the determination of \hc. First, time-delay measurements can be significantly less time-consuming for transients than for quasars because the light curve is not stochastic ~\citep{Liao2017,Goldstein2017}, and it generally has high amplitude and short duration. Second, transients usually do not suffer from microlensing effects as much as quasars. For example, ground-based GWs have much longer wavelengths than the lens scales, i.e., the Schwarzschild radii of stars \citep[in the wave optics limit, see][]{Liao2019} and thus are free of the microlensing magnification effect; microlensing of the SN can be circumvented by using the achromatic phase of the color curve~\citep{Goldstein2017,Huber2020}. 

In this paper, we explore another potential advantage of using lensed transients for time-delay cosmography over quasars: one can obtain a clean image of the lensed host galaxy, before or after the transient's appearance. Lensed arcs provide the most constraints on the lens model, and in the case of lensed quasars, they are typically out-shined by the bright point sources. In the case of transients, this source of contamination is eliminated.
This point has been noted before by~\citet{Holz2001,Goldstein2017} for lensed supernovae and by \citet{Liao2017} for gravitational waves. However, a quantitative estimation of the improvement of lens modelling with realistic simulations has not been made yet.

We use realistic simulations to quantify the improvement in modelling lensed transients over lensed quasars, based on the current quality of imaging data and the state-of-the-art modelling software.
The simulation and modelling approaches are very close to those adopted by the time-delay lens modelling challenge~\citep[TDLMC,][]{Ding2018, Ding2020}, in which the present capabilities of lens modelling codes are assessed using realistic mock datasets. 

This paper is structured as follows. In Section~\ref{sec:tdcos}, we briefly summarize the relevant aspects of strong lensing time-delay cosmography. In Section~\ref{sec:simulaiton}, we simulate the mock sample and then infer the lens models and \hc. The results with and without point sources are compared in Section~\ref{sec:result}. Conclusions are presented in Section~\ref{sec:conclusion}.

\section{Time delay cosmography}
\label{sec:tdcos}
The framework of strong lensing time-delay cosmography has been described before~\citep[see, e.g.,][]{Schneider1992, Blandford1992,Treu2016} and has been applied in previous studies to measure \hc~\citep[see, e.g.,][]{Suyu2017, Wong2017, Birrer2019, Shajib2020}. We refer the interested reader to these references for more details. Here we summarize only the main points for convenience of the reader and to establish the notation.

The time delay between any two lensed images (e.g., image $i$ and $j$) is given by:
\begin{eqnarray}\label{eq:td}
&\Delta t_{ij} = \frac{D_{\Delta t}}{c} \left[
\phi(\bm\theta_i)-
\phi(\bm\theta_j)
\right],
\end{eqnarray}
where $D_{\Delta t}$ is the so-called time-delay distance (see description later);
${\bm\theta}_i$ and ${\bm \theta}_j$ are the observed angular positions in the image plane~\citep{Refsdal1964, Shapiro1964}.
$\phi ({\bm \theta}_i)$ and $\phi ({\bm \theta}_j)$ are the corresponding Fermat potentials given by:
\begin{eqnarray}\label{eq:fermat}
& \phi ({\bm \theta}_i) \equiv \frac{1}{2}({\bm \theta}_i - {\bm \beta})^2 -
\psi({\bm \theta}_i),
\end{eqnarray}
where $\bm \beta$ is the unlensed source position and $\psi({\bm \theta}_i)$ is the scaled gravitational potential at the image position. The time delay caused by the first term, i.e., $\frac{1}{2}({\bm \theta}_i - {\bm \beta})^2 - \frac{1}{2}({\bm \theta}_j - {\bm \beta})^2$ is called the {\it geometric delay}, and that caused by the second term $\psi({\bm \theta}_i) - \psi({\bm \theta}_j)$ is called the {\it Shapiro delay}.
Modern lens modelling techniques (presented in Section~\ref{sec:lens_model}) are able to recover the lens model parameters and predict the Fermat potentials using high-resolution imaging.

The time-delay distance is a combination of three angular diameter distances:
\begin{eqnarray}
\label{eq:tdd}
& D_{\Delta t} \equiv (1+z_{\rm d}) \frac{D_{\rm d} D_{\rm s}}{ D_{\rm ds}},
\end{eqnarray}
where $D_{\rm d}$, $D_{\rm s}$ and $D_{\rm ds}$ are respectively the angular
distances from the observer to the deflector, from the observer to the source, and from the deflector to the source.
It is an absolute scale of the Universe as:
\begin{eqnarray}
\label{eq:inverto_H0}
& D_{\Delta t} \propto 1/H_0,
\end{eqnarray}
and it is weakly dependent
on other cosmological parameters, such as curvature and the matter density of the Universe.

A major limitation of the inference of $D_{\Delta t}$ is
the so-called mass-sheet transformation~\citep[MST, e.g.,][]{Falco1985} and its generalizations~\citep{Saha2000, Saha2006, Liesenborgs2012, Schneider2014, Birrer2016b, Wagner2018, Wertz2018}. 
The line-of-sight (LOS) structure produces an {\it external} MST which affects the inference of the time delay by:
\begin{eqnarray}\label{eq:k_ext_dt}
\Delta t^{\rm obs} = (1-\kappa_{\rm ext}) \Delta t^{\rm predict},
\end{eqnarray}
where $\Delta t^{\rm obs}$ is the observed time-delay and $\Delta t^{\rm predict}$ is the model predicted one.
As a consequence, the MST affects the inference of the time-delay distance by:
\begin{eqnarray}\label{eq:k_ext_tdd}
D_{\Delta t} = \frac{1}{(1-\kappa_{\rm ext})} D_{\Delta t}',
\end{eqnarray}
where $D_{\Delta t}'$ is the derived time-delay distance without considering the external MST effect and the $D_{\Delta t}$ is the true value. 
In practice, this external convergence factor $\kappa_{\rm ext}$, which accounts for the
contribution of all the mass along the line of sight, is inferred 
independently from the lens modelling process. By comparing the relative numbers of galaxies weighted by physically relevant 
quantities in terms of their distances to the lens and their own stellar masses and redshifts, the probability distribution of $\kappa_{\rm ext}$ is estimated using numerical simulations based on similar statistical properties~\citep{Rusu2017}. Alternatively, weak lensing analysis can also estimate the external convergence~\citep{Tihhonova2018}.
 
In addition to the external MST, there is an {\it internal} MST. Mathematically, for every distribution of mass that matches the lensed images one can obtain a family of solutions by applying a rescaling that leaves the lens equation unchanged. The fundamental question is whether this rescaling is physical or not. 
This degeneracy can be broken in two ways: i) by applying theoretical priors, e.g., from galaxy simulations if they are of sufficient quality; ii) by applying non-lensing data, such as, for example, stellar kinematics \citep{TreuKoopmans2002,Koopmans2003}, as implemented in the hierarchical Bayesian approach developed by~\citet{Birrer2020}.

\section{Data simulation and modelling test}
\label{sec:simulaiton}
\begin{figure*}
    \centering
    \includegraphics[trim = 50mm 50mm 40mm 50mm, clip,width=0.7\textwidth]{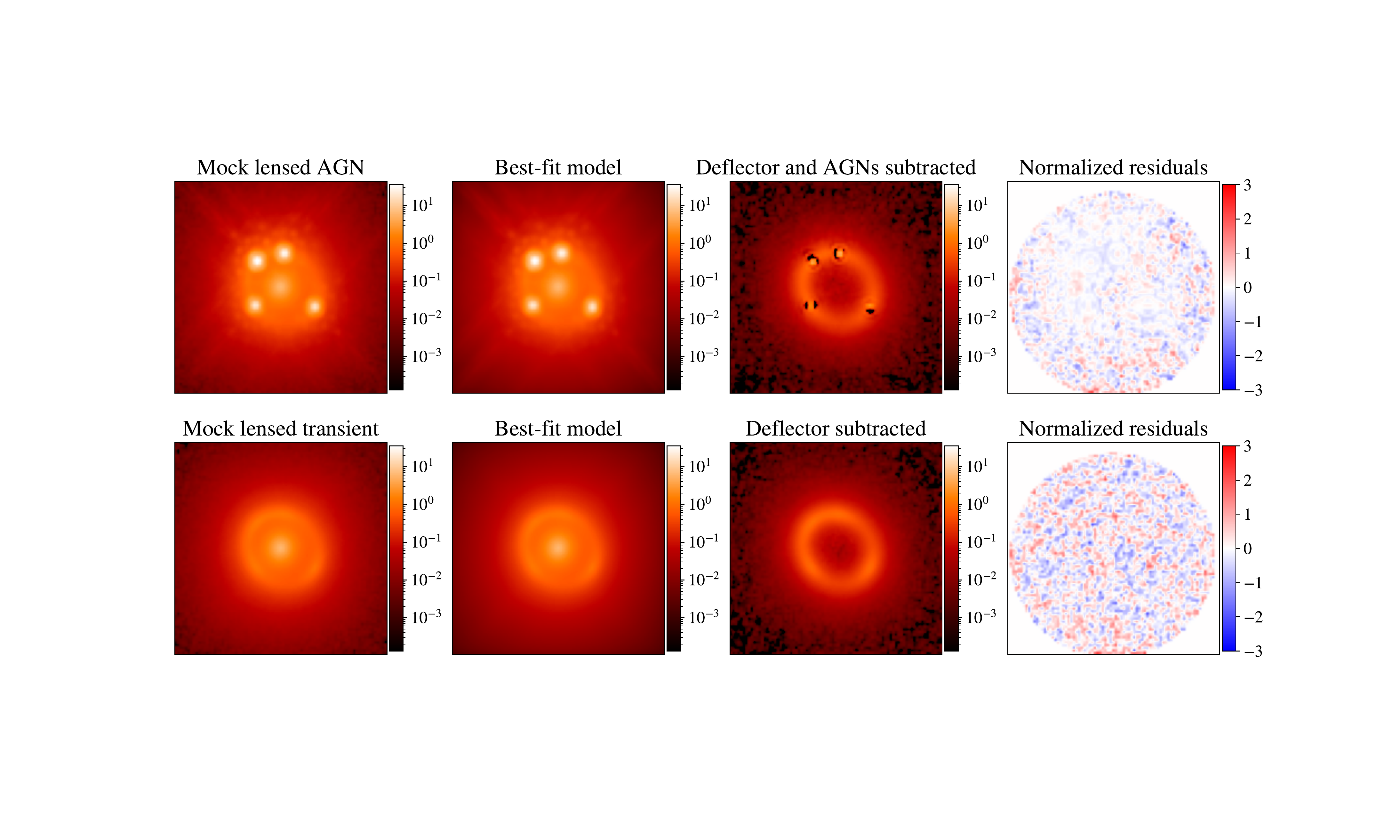}
    \caption{One example pair from the simulated lenses (48 pairs in total) for lensed quasar case (top) and transient non-AGN case (bottom). Figures from left to right: 1) simulated \hst\ F160W images of the lensing systems, using a logarithmic color scale with the same stretch. 2) best-fit model. 3) lensed arcs after removing the lights from the deflectors (and point sources for lensed AGN case), based on the best-fit model. Note that the arc surface brightness is the key to infer the lens model parameters. As shown in the upper-third panel, the mismatch between the PSF model and the actual AGN signal creates large residual patterns, which affect the model inference. 4) normalized residual maps based on the best-fit models. For the lensed AGN case, the fitting result is plotted based on a 50\% PSF uncertainty level assumption.
    }
    \label{fig:fitting_comparison}
\end{figure*}

One of the major challenges in lensed quasar image modelling is the brightness of the AGN images --- they are unresolved and dominate the surface brightness of host galaxy arcs near the images. The common practice is to 
start with a nearby star in the field of view and used it as a point spread function (PSF) model to fit the data. The bright point sources are modelled as scaled PSFs. The inevitable mismatch between the star PSF and the AGN PSF (due to, e.g., under-sampling, pixel responsivity, intrinsic color, spatial variations) generally leaves significant residuals at the lensed AGN positions. Mitigation strategies include boosting the noise level 
in the AGN central regions~\citep{Suyu2013, Birrer2016} to account for the PSF error, and/or use an iterative scheme~\citep{Chen2016, Birrer2019} to improve the PSF estimate during the lens model by taking advantage of the presence of 4 AGN images. Even with a perfect PSF, a large number of photons from the bright AGN images significantly increases the Poisson noise affecting the underlying extended components, and thus affecting the error budget of the lens model inference.
In any case, the presence of bright point sources degrades the lensed arc information and reduces the precision of the inferred lens model.

The goal of this work is to
quantitatively
evaluate the effect of the AGN light contamination on the lens modelling. To this end, we use simulations and build two samples to perform a controlled experiment.
The first sample consists of 48 mock lensed quasars (i.e., lensed sources are as galaxy+AGN) based on \hst\ images. The data quality mimics that achieved on real observation by the TDCOSMO collaboration. We generate a control sample using the same pipeline with the 
same parameter values, i.e., the lens galaxies and host galaxies are the same between the two samples.
The only difference is that the AGN is not added to the source galaxy in the control sample. We then utilize the lens modelling software \textsc{lenstronomy}\footnote{\url{https://github.com/sibirrer/lenstronomy}}~\citep{lenstronomy} with typical modelling strategies to model the two samples. By comparing their results, we can quantify how much the precision of the inferred lens parameters and \hc\ improve once the bright point sources disappear, as one would do in the case of transient point sources.

Of course, the disappearance of the bright point sources would also facilitate the determination of the stellar kinematics of the deflector, providing an additional benefit. In this paper, we focus on the improvement of lens modelling and leave the quantification of stellar kinematics improvement in regard to the internal MST and radial mass profile for future work.

\subsection{Mock data simulations}\label{sec:mock}

The mock images are generated based on the pipeline introduced by~\citet{Ding2017, Ding2017b}. This pipeline was also used to simulate the sample in the time-delay lens modelling challenge \citep[TDLMC,][]{Ding2018} and in TDCOSMO collaboration \citep{Millon2020}. A brief summary of the simulation is described below, and we refer to~\citet{Ding2017} for more details. 

The simulation consists of the following steps: 1) determining the models and parameters of the lensing system (i.e., light and mass profile); 2) using the lens equation to calculate the surface brightness and project it on a pixelated frame (the initial pixel resolution is higher than \textit{HST} WFC3's by a factor of four); 3) using the high-resolution PSF model to convolve the image (for lensed quasars, the AGN images is added in this step); 4) rebinning the images down to \hst's resolution according to the different dithering patterns; 5) adding noise; 6) drizzling images to get the final science image. Steps 2 and 3 are performed using \textsc{lenstronomy}.

The light distribution of the deflector galaxy is modelled on the image plane, and the background source galaxy is modelled on the source plane.
For the light profile of the galaxy (for both deflector galaxy and source galaxy), we adopt a standard S\'ersic model, which is parameterized as:
\begin{eqnarray}
   \label{eq:sersic}
   &I(R) = A \exp\left[-k\left(\left(\frac{R}{R_{\mathrm{eff}}}\right)^{1/n}-1\right)\right] ,\\
   &R(x,y,q) = \sqrt{qx^2+y^2/q},
\end{eqnarray}
where $A$ is the amplitude and S\'ersic index $n$ controls the shape of the radial
surface brightness profile: larger $n$ corresponds to a steeper
inner profile and a highly extended outer wing. 
 $k$ is a constant which
depends on $n$ so as to ensure that the isophote at $R=R_{\mathrm{eff}}$
encloses half of the total light~\citep{C+B99}, and
$q$ denotes the axis ratio. The light of the 
AGN is described by a point source in the image plane.

The mass of the deflector is described by an elliptical power-law model, whose surface mass density is given by:
\begin{eqnarray}
 \label{massmodel}
 \Sigma(x,y)=\Sigma_{\rm cr}\frac{3-\gamma}{2}\left(\frac{\sqrt{q_{\rm m} x^{2}+y^{2}/q_{\rm m}}}{R_{\rm E}}\right)^{1-\gamma},
\end{eqnarray}
where $q_{\rm m}$ describes the projected axis ratio.
The so-called Einstein radius $R_{\rm{E}}$ encloses a mean surface density equal to $\Sigma_{\rm cr}$ when $q_{\rm m}=1$ (i.e. spherical limit).

The exponent $\gamma$ is the absolute value of the logarithmic slope of the power-law profile,
for massive elliptical galaxies $\gamma \approx2$~\citep{T+K02a,T+K04,Koo++09, Shajib2020b}.
We refer the reader to the reviews by~\citet{Sch06, Bar10, Tre10} for more details.
An external shear component is also added. 

The input parameters of the sample are randomly generated based on the distributions listed in Table~1 of TDLMC-II~\citep{Ding2020}, except that the external $\kappa_{\rm ext}$ is not considered for simplicity (it is irrelevant for the argument presented here). The position of the source galaxy is chosen so as to obtain quadruple images of the point sources.

We use {\sc tinytim}~\citep{Krist2011} to generate the PSF, used both for convolution of extended light and for representing the images of the AGN. 
For each system, we rebin the high-resolution image eight times at eight different frame positions to reproduce the dither process in the real observation~\citep[see Fig.~2 in][]{Ding2017}.
Noise is then added to the mocks, which consists of Gaussian background noise and Poisson noise. The noise level is based on the realistic \hst\ condition with total exposure time assumed to be $8\times 1200 {\rm\ s} = 9600$ s. Finally, we use \textsc{multidrizzle}\footnote{\url{https://www.stsci.edu/koekemoe/multidrizzle/}} and co-add the images and PSF to increase the pixel resolution from 0\farcs{13} to 0\farcs{08}, ending with cutouts of  $99\times99$ pixels.

The same pipeline with identical setting is used to generate the lensed transient images, of course with the exception of the addition of the point images.

The mock time-delay values are calculated based on the lens parameters in a standard flat $\Lambda$CDM model, with $\Omega_{\rm m} = 0.27$ and $\Omega_\Lambda$ = 0.73. The true \hc\ value is randomly assumed to be 73.9~km~s$^{-1}$~Mpc$^{-1}$, for convenience, although we note that our results are independent of the adopted value. For the time-delay uncertainty, we assume a non-biased error with the root-mean-square (rms) level as either 1\% level or 0.25 days, depends on which one is larger, representing the best case scenario for current datasets.

In Fig.~\ref{fig:fitting_comparison} (left), we illustrate one example pair of the mock lensed quasar and lensed transient images.  Since the two members of each pair share the same parameters, the time delays are the same.

\subsection{Lens model inference}
\label{sec:lens_model}
We adopt a standard approach to infer the lens model and combine it with the time-delay information to derive \hc. We follow common practice and define a mask region to bracket the pixels with sufficient signal-to-noise level (signal to background rms ratio $> 5$ in this work) and constrain the model parameters. 
To efficiently search for a maximum in the likelihood space, we use \textsc{lenstronomy} and first apply a  Particle Swarm Optimizer (PSO) to optimize the model~\citep{Kennedy1995, Birrer2015}. We run the PSO four times and then utilize a Markov Chain Monte Carlo (MCMC) sampler algorithm~\citep[\textsc{emcee};][]{Foreman-Mackey2013} to derive the best-fit parameters and their uncertainties.
All the free parameters used in the simulations are considered in the fitting, which are from the S\'ersic light profiles (7 parameters, see Eq.~\ref{eq:sersic}) for lens and source galaxy, elliptical power-law model (6 parameters, see Eq.~\ref{massmodel}), external shear (2 parameters), and lensed point source positions. For lensed AGN cases, the lensed AGN flux is also a free parameter.
To check that the global minimization has been obtained for each system, we run the fitting three times independently --- each time randomizing the input particle positions --- and find that consistent results are obtained every time with our adopted numerical settings.

For the lensed AGN case, our simulations are designed to include the fact that the true PSF is unknown in real data. 
Note that the PSF model used in the fitting is initially the same as the AGN light distribution, which is also produced by {\sc tinytim}. However, in the third step of simulation (see Section~\ref{sec:mock}), interpolation is performed to add the PSF at an arbitrary position on the pixel as the lensed AGN image with pixellization effects.
Moreover, an additional distortion is introduced after the drizzling when the light distribution has steep gradients. These effects could produce a mismatch between PSF and AGN that up to 50\% error in the central region.
When modelling the lensed AGN images, we account for this mismatch by boosting the PSF uncertainty level. In \textsc{lenstronomy}, the boosted PSF uncertainties are implemented as a fraction of the AGN surface brightness and added to the final noise map. For example, a 10\% boost in PSF uncertainty means that the pixels close to the lensed point sources have their statistical error increased by summing in quadrature 10\% of the counts in the PSF at that point. The PSF is a significant source of noise and therefore this step is essential for obtaining realistic error estimates on \hc.
To find the correct noise level for our setup and avoid overfitting\footnote{In principle, we aim at inferring the \hc\ with a proper error bar level to describe the random scatter of the best-fit values.}, we consider three levels of boosted PSF uncertainty (i.e., 10\%, 30\%, and 50\%).
For the lensed transient case, since the AGN images are not added, the issue of PSF mismatch does not exist and therefore this step is not necessary.

The positions of the point source in the image plane and source plane are needed to calculate the time delays (i.e., $\bm \theta$ and $\bm \beta$ in Eq.\ref{eq:td} and~\ref{eq:fermat}). However, for our transient case, the point source is not visible during the time of exposure, and thus the lensed point source positions in the image plane (hence its position in the source plane) cannot be inferred to sufficient precision using the lensed host galaxy images alone. In practice, one will have to infer the astrometry of the transient source $\bm \theta$ while it is visible and then apply it to the images obtained without transient. This step will require registering the images and correcting for distortion at sufficient level of precision \citep[see][for astrometric requirements in such a process]{Birrer2019}. We assume for this work that this step has been taken care of, and that the precision of the lensed transient position is the same as that of the AGN positions ($\sim$0.004 arcsec).

\section{Results}
\label{sec:result}
Based on the modelling approach introduced in the previous section, we illustrate the best-fit models for a pair of twin systems in Fig.~\ref{fig:fitting_comparison} together with the corresponding lensed arcs remnants (i.e., removing light from deflector and point sources) and the normalized residual maps. For the lensed quasars case, the mismatch of the PSFs can be seen clearly at the AGN central regions in the top-third panel.
In the remainder of this section, we compare the inferences of the lens models and \hc\ between the lensed AGN and transient. 

\subsection{Improvement in terms of \hc}
\begin{figure}
\centering
\hspace*{-1cm}    
{\includegraphics[height=0.17\textwidth]{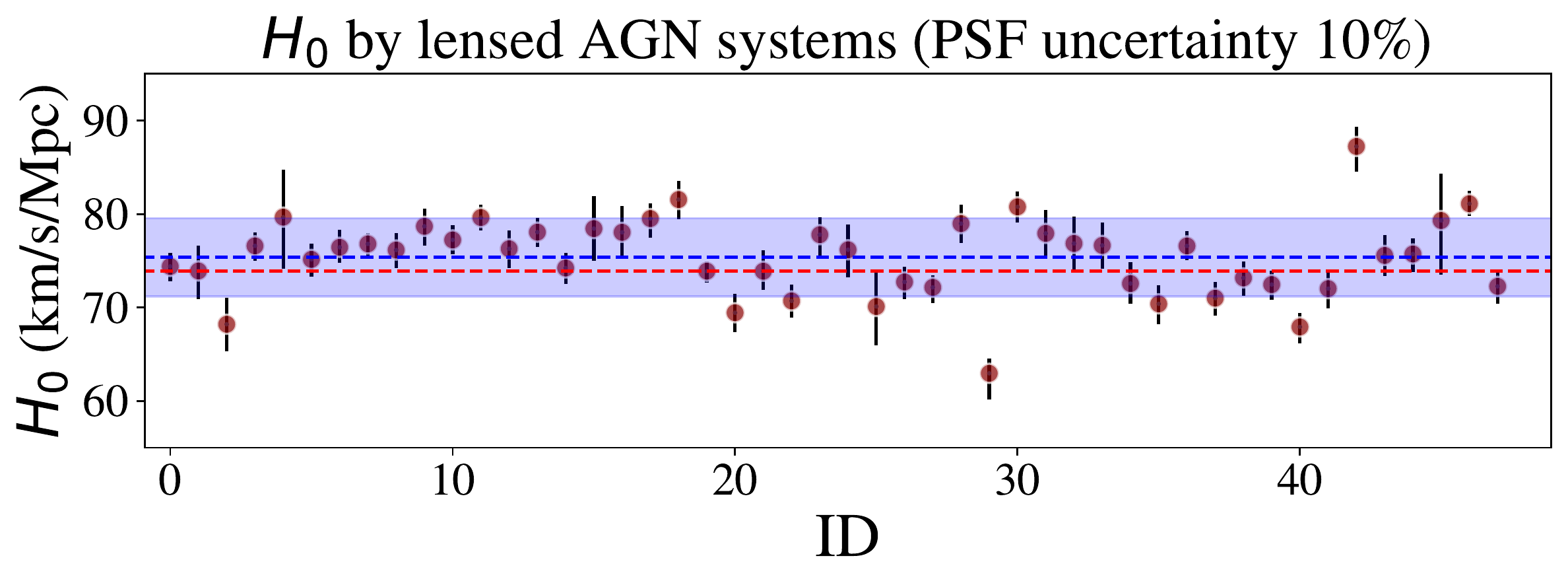}}\\
\hspace*{-1cm}   
{\includegraphics[height=0.17\textwidth]{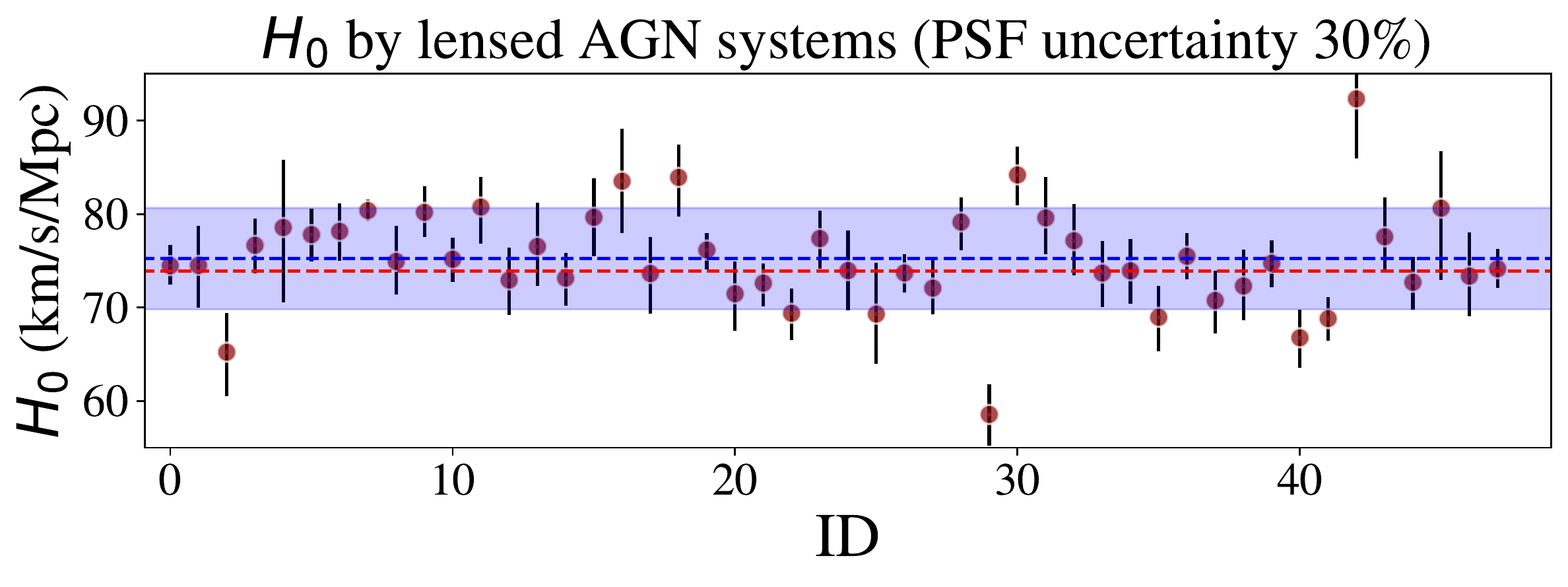}}\\
\hspace*{-1cm}
{\includegraphics[height=0.17\textwidth]{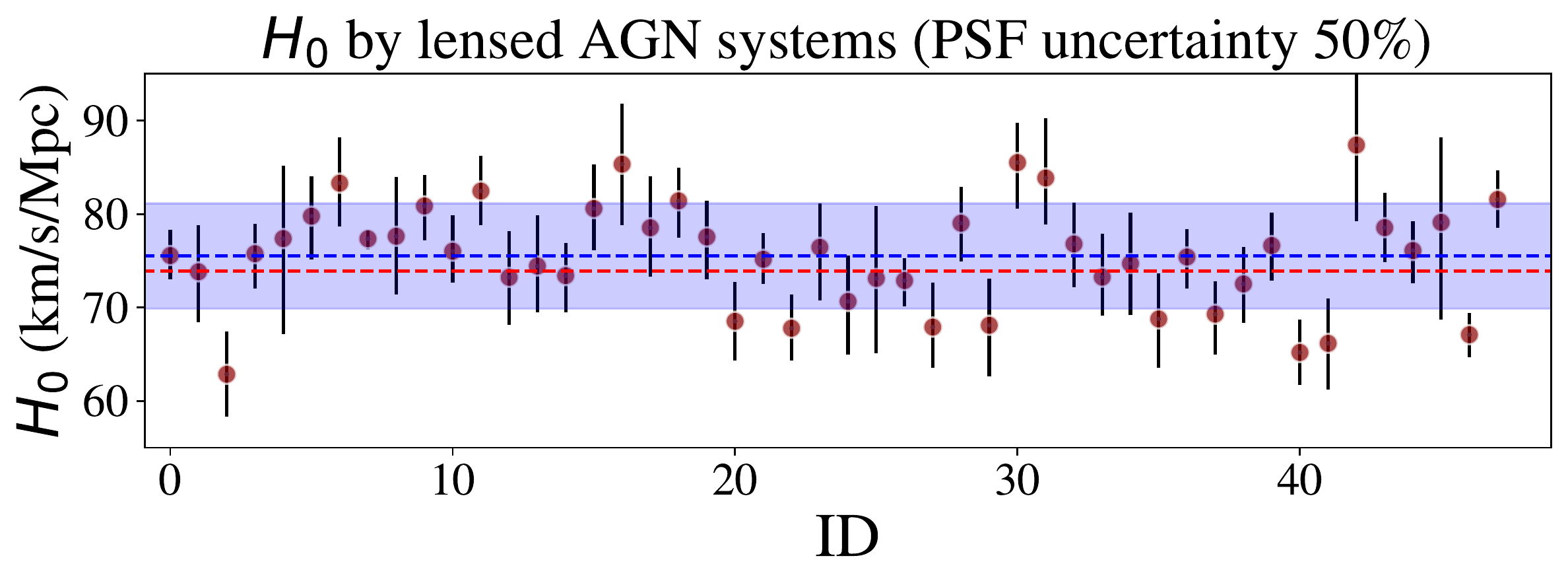}}\\
\hspace*{-1cm}   
{\includegraphics[height=0.17\textwidth]{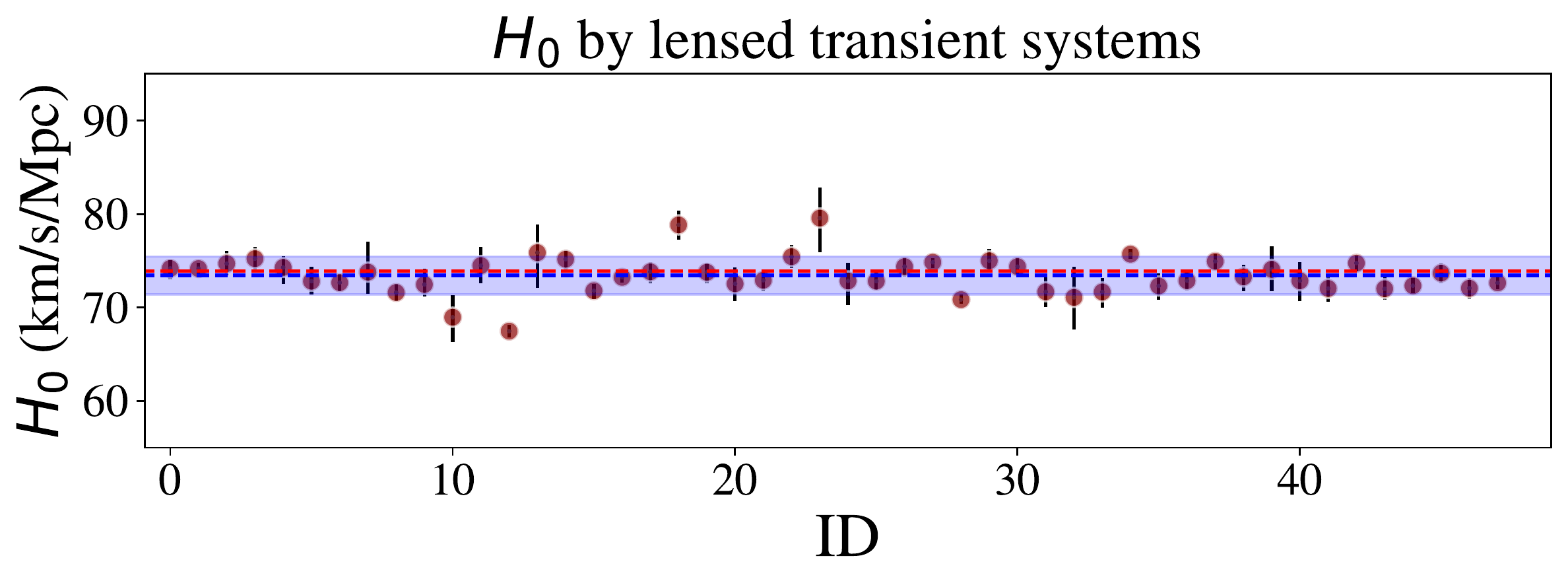}}\\
\caption{\label{fig:h0_distribution} 
Inferred \hc\ for 48 lensed AGN systems (top 3 panels) and 48 lensed transient systems (bottom panel). 
For lensed AGN systems, three different PSF uncertainty levels are assumed (i.e., 10\%, 30\% and 50\%) in the lens modelling process, respectively.
The red dashed line indicates the true \hc\ value; the blue dashed line and blue region are the mean and standard deviation of the best-fit \hc\ measurements.
For our fiducial case (corresponding to 50\% uncertainties on the PSF), the average precision of \hc\ per system is 7.4\% for the lensed AGN systems. For the transients, the precision is improved to 2.7\%.
}
\end{figure}

We present the inferences of \hc\ together with the 1-$\sigma$ uncertainty levels as error bars for the 48 systems for lensed AGNs and lensed transients in Fig.~\ref{fig:h0_distribution}. As described in Section~\ref{sec:lens_model}, for the lensed AGN sample, we explored three different PSF uncertainty levels (i.e., 10\%, 30\%, and 50\%) to infer the lens model, and the resulting \hc\ based on the three different settings are presented in the top three panels of Fig.~\ref{fig:h0_distribution}, respectively. 
Based on the best-fit values, the mean and standard deviation  of \hc\ are $75.4\pm4.2$, $75.2\pm5.4$, and $75.5\pm5.6$ (in a unit of~\hcunit
) for 10\%, 30\%, and 50\% PSF uncertainty levels, respectively, while the true value of \hc\ is 73.9~\hcunit.
As expected, increasing the PSF uncertainty level increases the uncertainty on \hc, while the best-fit \hc\ values 
remain unchanged in general.
Among these three results, we find that the error bars are underestimated for PSF uncertainties at the 10\% and 30\% level, while the standard deviation value of the best-fits is consistent with the estimated errors when the PSF uncertainty level is set at 50\%. We conclude that this choice provides the most realistic accounting of this source of uncertainty. Thus, we take the 50\% PSF uncertainty level as our fiducial case in the rest of the paper, and consider the corresponding standard deviation value (i.e., $\Delta H_0 =$ 5.6~\hcunit) as the inferred precision level of \hc\ for the lensed AGN case. We note that this precision level is consistent with the performance of the Student-T team in the TDLMC~\citep{Ding2020}, who adopted a similar approach like this work to a similar mock lensed-AGN data.

For the lensed transient (non-AGN) case, the realization result of the \hc\ based on the 48 system is improved to $73.6\pm1.9$~\hcunit. 
The main result of this work is that the precision of \hc\ inferred from the lensed transients is improved by a factor of 2.9 with respect to the lensed AGNs.

\subsection{Improvement in terms of lens model}
Having found that the lensed transient can yield higher precision \hc\ than lensed quasars, we expect this is due to the fact that the lens models in the lensed transient have been derived at higher precision. We also expect to see the tight relation of the inference between the lens model parameters and \hc.
As a sanity check, we now compare in detail the precision of the inferred lens model parameters to confirm the origin of the uncertainty on \hc.

\begin{figure*}
\centering
\begin{tabular}{c c}
{\includegraphics[height=0.4\textwidth]{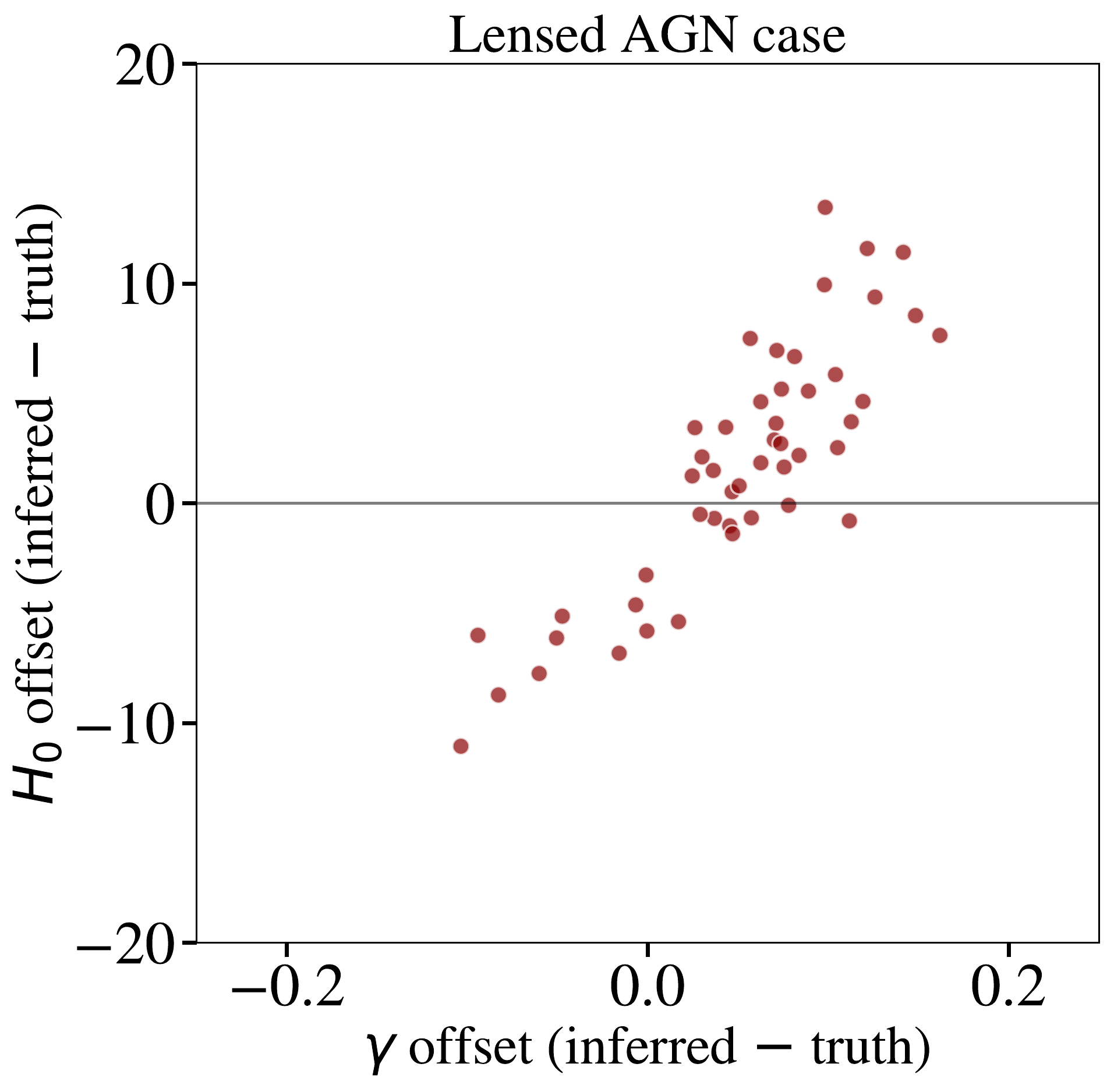}}&
{\includegraphics[height=0.4\textwidth]{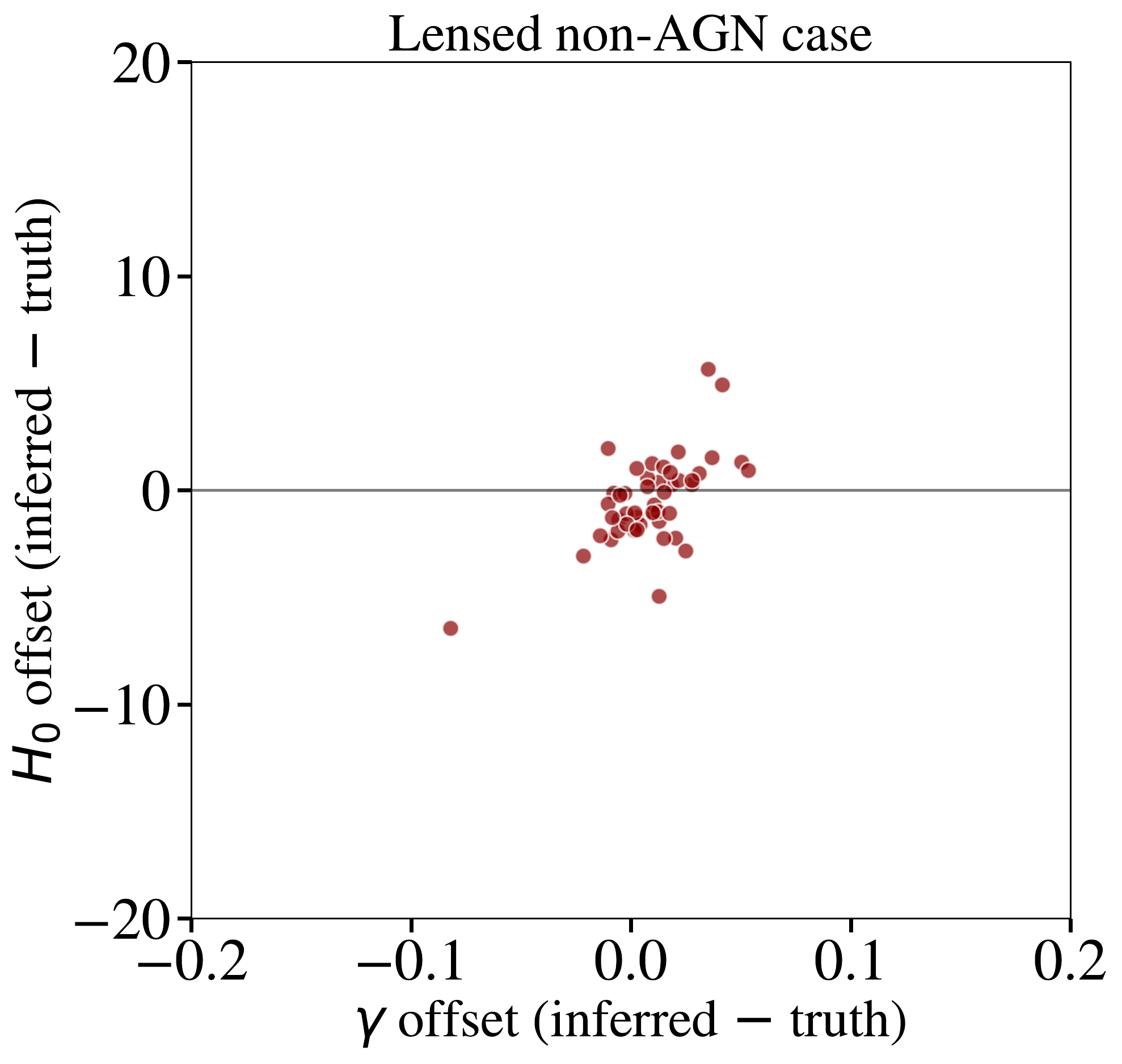}}\\
\end{tabular}
\caption{\label{fig:bias_result} 
Offset distributions of the inferred best-fit \hc\ as a function of inferred best-fit $\gamma$, based on 48 simulated systems, for lensed AGN case (left) and non-AGN case (right).
For the lensed AGN case, the result is based on the assumption of $50\%$ PSF uncertainty level.
}
\end{figure*}

It is well known that the slope $\gamma$ in Eq.~\ref{massmodel} of total mass density profile of the deflector is the dominant factor in determining \hc\ amongst the lens model parameters \citep{Witt2000}. Therefore, we plot the distribution of the offset value (i.e., the difference between inferred best-fit and true value) of \hc\ as a function of the $\gamma$ offset in Fig.~\ref{fig:bias_result} for lensed AGN and lensed transient, respectively. As expected, the $\gamma$ offset distribution shows a larger scatter and a strong correlation with the \hc\ distribution for the lensed AGN case, see Fig.~\ref{fig:bias_result} (left). This is consistent with the previous finding by \citet{Witt2000} that the inferred \hc\ scales approximately as $H_0 \propto (\gamma - 1)$. Of course, there are other sources of uncertainty in our inference simulation (for example, uncertainty in the time delay), which introduce additional scatter in the \hc\ and $\gamma$ correlation. Therefore, this trend is less pronounced in the lensed transient case, where the $\gamma$ offset scatter is smaller, see Fig.~\ref{fig:bias_result}, (right). We calculate the mean and standard deviation of the $\gamma$ offset. The results are $0.050\pm0.061$ and $0.013\pm0.015$ for lensed AGN system and non-AGN system, respectively. The lens model's precision, in terms of $\gamma$, improves by a factor of 4.1 once the AGN images are removed.

\begin{figure*}
\centering
\begin{tabular}{c c}
{\includegraphics[height=0.4\textwidth]{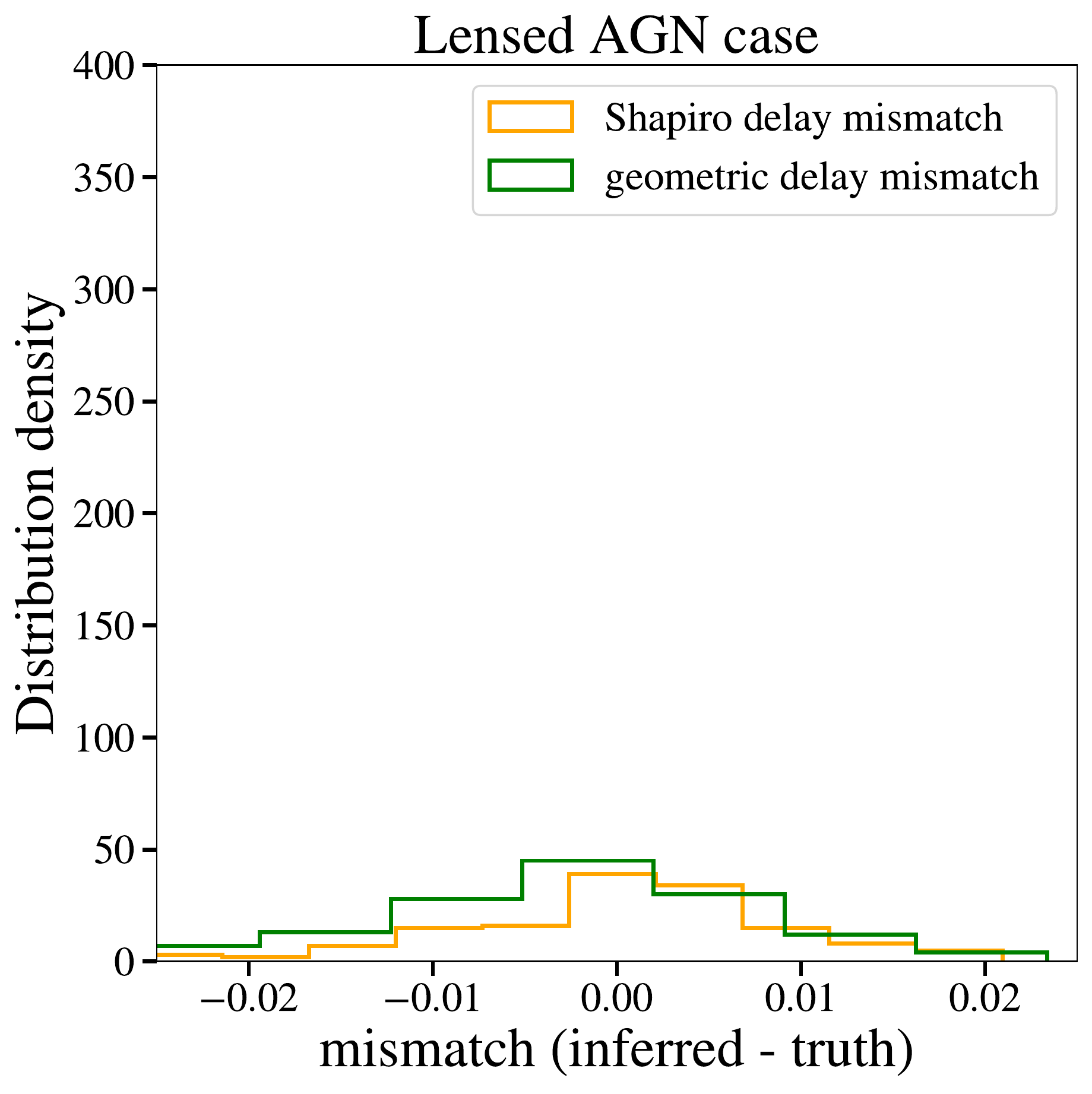}}&
{\includegraphics[height=0.4\textwidth]{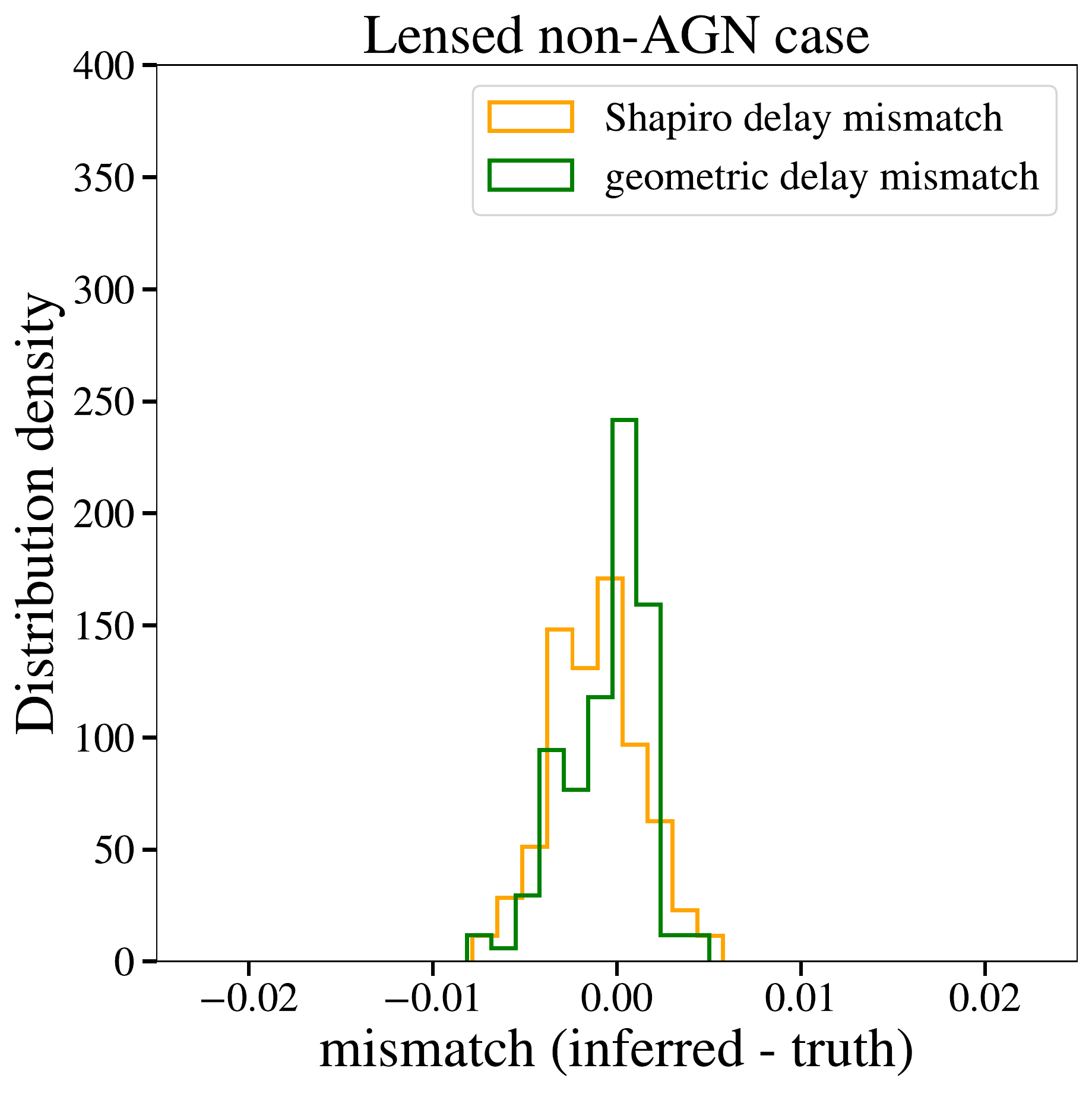}}\\
\end{tabular}
\caption{\label{fig:Fermat_bias} 
Histograms of the inferred best-fit Shapiro potential delay [i.e.,  
difference between the inferred and the truth
of the $\psi({\bm \theta}_i) - \psi({\bm \theta}_j)$ value] and geometric potential delay [i.e., 
difference between the inferred and the truth
of the $\frac{1}{2}({\bm \theta}_i - {\bm \beta})^2 - \frac{1}{2}({\bm \theta}_j - {\bm \beta})^2$ value] between each two lensed images for lensed AGN case (left) and non-AGN case (right).}
\end{figure*}

The dependency of \hc\ on lens model parameters has been studied analytically before \citep[e.g.,][]{Witt2000, Wucknitz2002, Kochanek2002}. 
The lens model defines the Fermat potential and thus determines the \hc\ through Eq.~\ref{eq:td} and Eq.~\ref{eq:inverto_H0}. Therefore, the higher precision of \hc\ implies a better constrained Fermat potential, which is composed of two terms involving the geometric delay and the Shapiro delay (see Eq.~\ref{eq:fermat}).
As another sanity check, we calculate the offset of the Shapiro delay difference and the geometric delay difference for the 48 systems\footnote{For a lensed quadruple system with lensed images at ABCD, we calculate the inferred Shapiro delay difference and geometric delay difference between three pairs, i.e., A-B, A-C, A-D, and compare them to the truth value. For 48 lensed systems, there are 144 pairs calculated and plotted in the histogram in Fig.~\ref{fig:Fermat_bias}.} to understand how much they improve, respectively. We plot the offset distribution in Fig.~\ref{fig:Fermat_bias}.
As expected, we find that the distributions for the lensed AGN sample have a larger scatter for both the Shapiro delay and the geometric delay. 
For the lensed AGN systems, the mean and standard deviation results are $(0.44\pm8.68) \times10^{-3}$ and  $(-3.07\pm11.53)\times10^{-3}$ (in a unit of arcsec$^2$) for the Shapiro delay mismatch and the geometric delay mismatch, respectively. For lensed non-AGN case, the results are $(-0.56\pm2.28)\times10^{-3}$ and $(-1.08\pm2.44)\times10^{-3}$ (in a unit of arcsec$^2$), respectively. By comparing the standard derivations, we find that the lensed transient sample has higher precision by a factor of 3.8 and 4.7 on the Shapiro delay and the geometric delay, respectively.

These results support our conclusion that the precision of \hc\ is improved in the lensed transient systems because the lens model parameters can be constrained more tightly.

\section{Conclusions}
\label{sec:conclusion}
Strongly lensed transients are expected to be readily detected and observed with upcoming facilities in the near future.
Time-delay cosmography applied to these systems will be a powerful complement to that based on lensed quasars.
Compared with lensed quasars, the lensed transients can provide both time-delay measurements and lens model inferences more precisely.
For the latter, lensed transient systems can be observed (before or after the transient) with complete and clear host images without contamination from bright point-source images (i.e., AGN). 

We performed the first quantitative study on the improvements of lens modelling in lensed transient systems by carrying out realistic simulations based on state-of-the-art observations and algorithms. Mimicking standard \hst\ imaging conditions, we simulated a sample of 48 lensed AGN systems. Using the same pipeline and the same parameter values, we simulated 48 paired lensed transient systems that do not have AGNs in the host. We then adopted typical modelling strategies to estimate \hc\ from the lens systems and made a direct comparison of the inferences between the two samples.

We showed that, compared with lensed AGN systems, the inferred precision of \hc\ is improved by a factor of 2.9 once the AGN is removed in the lensed transient systems. As a sanity check, we found that \hc\ is strongly related to the logarithmic slope $\gamma$ of the mass profile. In lensed transient systems, the precision of $\gamma$ is improved by a factor of 4.1 with respect to non-transient lenses. We conclude that in transient systems, higher precision of lens model parameters can be obtained at fixed observational conditions, which in turn improves the overall precision of \hc.

We note that this improvement level is also determined by the observational conditions. 
In our simulation exercise, a particular set of instruments is assumed, i.e., \hst\ WFC3/F160W. However, we expect that the level of improvement depends on the data quality and imaging resolution. For example, at a lower angular resolution, the bright point sources would affect the results more, thus making transients more advantageous than reported here.
Conversely, at a higher angular resolution than \hst, for example, with the \textit{James Webb Space Telescope} or advanced adaptive optics from the ground, we expect that the effects of the point source contamination on the lens models will be less than those reported in this work.

Besides the improvement of the lens modelling discussed in this work, the precision of the \hc\ can be even further improved considering the other aspects of the lensed transients. For example, we stress that the kinematics of the lens galaxy can be measured more easily for the transient case, since one can avoid the AGN light contamination. 
This is potentially a very important advantage considering that improved measurements of the stellar kinematics are crucial to breaking the mass-sheet degeneracy~\citep{Birrer2020}.
Moreover, for lensed type Ia SNe as standard candles and GWs as standard sirens, we can directly measure magnification factors and use those to help break the mass-sheet degeneracy.

\section*{Acknowledgments}
KL was supported by the National Natural Science Foundation of China (NSFC) No. 11973034. 
This research was supported by the U.S. Department of Energy (DOE) Office of Science Distinguished Scientist Fellow Program. 
TT acknowledges support by a Packard Research Fellowship, and by NSF through grant AST-1906976 "Collaborative Research: Toward a 1\% measurement of the Hubble Constant with gravitational time delays".
This work was supported by World Premier International Research Center Initiative (WPI), MEXT, Japan.

\section*{Data availability}
The simulated data underlying this article are available in \url{https://github.com/dartoon/publication/tree/main/simulation_48_paired_lens}.

\bibliographystyle{mnras}

\begin{thebibliography}{}
\makeatletter
\relax
\def\mn@urlcharsother{\let\do\@makeother \do\$\do\&\do\#\do\^\do\_\do\%\do\~}
\def\mn@doi{\begingroup\mn@urlcharsother \@ifnextchar [ {\mn@doi@}
  {\mn@doi@[]}}
\def\mn@doi@[#1]#2{\def\@tempa{#1}\ifx\@tempa\@empty \href
  {http://dx.doi.org/#2} {doi:#2}\else \href {http://dx.doi.org/#2} {#1}\fi
  \endgroup}
\def\mn@eprint#1#2{\mn@eprint@#1:#2::\@nil}
\def\mn@eprint@arXiv#1{\href {http://arxiv.org/abs/#1} {{\tt arXiv:#1}}}
\def\mn@eprint@dblp#1{\href {http://dblp.uni-trier.de/rec/bibtex/#1.xml}
  {dblp:#1}}
\def\mn@eprint@#1:#2:#3:#4\@nil{\def\@tempa {#1}\def\@tempb {#2}\def\@tempc
  {#3}\ifx \@tempc \@empty \let \@tempc \@tempb \let \@tempb \@tempa \fi \ifx
  \@tempb \@empty \def\@tempb {arXiv}\fi \@ifundefined
  {mn@eprint@\@tempb}{\@tempb:\@tempc}{\expandafter \expandafter \csname
  mn@eprint@\@tempb\endcsname \expandafter{\@tempc}}}

\bibitem[\protect\citeauthoryear{{Abbott} \& {Virgo Collaboration}}{{Abbott} \&
  {Virgo Collaboration}}{2019}]{GWTC1}
{Abbott} B.~P.~L.,  {Virgo Collaboration} 2019, \mn@doi [Physical Review X]
  {10.1103/PhysRevX.9.031040}, \href
  {https://ui.adsabs.harvard.edu/abs/2019PhRvX...9c1040A} {9, 031040}

\bibitem[\protect\citeauthoryear{{Aiola} et~al.,}{{Aiola}
  et~al.}{2020}]{ACT2020}
{Aiola} S.,  et~al., 2020, \mn@doi [\jcap] {10.1088/1475-7516/2020/12/047},
  \href {https://ui.adsabs.harvard.edu/abs/2020JCAP...12..047A} {2020, 047}

\bibitem[\protect\citeauthoryear{{Bartelmann}}{{Bartelmann}}{2010}]{Bar10}
{Bartelmann} M.,  2010, \mn@doi [Classical and Quantum Gravity]
  {10.1088/0264-9381/27/23/233001}, \href
  {http://adsabs.harvard.edu/abs/2010CQGra..27w3001B} {27, 233001}

\bibitem[\protect\citeauthoryear{{Biesiada}, {Ding}, {Pi{\'o}rkowska}  \&
  {Zhu}}{{Biesiada} et~al.}{2014}]{Biesiada2014}
{Biesiada} M.,  {Ding} X.,  {Pi{\'o}rkowska} A.,   {Zhu} Z.-H.,  2014, \mn@doi
  [\jcap] {10.1088/1475-7516/2014/10/080}, \href
  {https://ui.adsabs.harvard.edu/abs/2014JCAP...10..080B} {2014, 080}

\bibitem[\protect\citeauthoryear{{Birrer} \& {Amara}}{{Birrer} \&
  {Amara}}{2018}]{lenstronomy}
{Birrer} S.,  {Amara} A.,  2018, \mn@doi [Physics of the Dark Universe]
  {10.1016/j.dark.2018.11.002}, \href
  {http://adsabs.harvard.edu/abs/2018PDU....22..189B} {22, 189}

\bibitem[\protect\citeauthoryear{{Birrer} \& {Treu}}{{Birrer} \&
  {Treu}}{2020}]{Birrer2020a}
{Birrer} S.,  {Treu} T.,  2020, arXiv e-prints, \href
  {https://ui.adsabs.harvard.edu/abs/2020arXiv200806157B} {p. arXiv:2008.06157}

\bibitem[\protect\citeauthoryear{{Birrer}, {Amara}  \& {Refregier}}{{Birrer}
  et~al.}{2015}]{Birrer2015}
{Birrer} S.,  {Amara} A.,   {Refregier} A.,  2015, \mn@doi [\apj]
  {10.1088/0004-637X/813/2/102}, \href
  {https://ui.adsabs.harvard.edu/abs/2015ApJ...813..102B} {813, 102}

\bibitem[\protect\citeauthoryear{{Birrer}, {Amara}  \& {Refregier}}{{Birrer}
  et~al.}{2016a}]{Birrer2016b}
{Birrer} S.,  {Amara} A.,   {Refregier} A.,  2016a, \mn@doi [\jcap]
  {10.1088/1475-7516/2016/08/020}, \href
  {https://ui.adsabs.harvard.edu/abs/2016JCAP...08..020B} {2016, 020}

\bibitem[\protect\citeauthoryear{{Birrer}, {Amara}  \& {Refregier}}{{Birrer}
  et~al.}{2016b}]{Birrer2016}
{Birrer} S.,  {Amara} A.,   {Refregier} A.,  2016b, \mn@doi [\jcap]
  {10.1088/1475-7516/2016/08/020}, \href
  {https://ui.adsabs.harvard.edu/abs/2016JCAP...08..020B} {2016, 020}

\bibitem[\protect\citeauthoryear{{Birrer} et~al.,}{{Birrer}
  et~al.}{2019}]{Birrer2019}
{Birrer} S.,  et~al., 2019, \mn@doi [\mnras] {10.1093/mnras/stz200}, \href
  {https://ui.adsabs.harvard.edu/abs/2019MNRAS.484.4726B} {484, 4726}

\bibitem[\protect\citeauthoryear{{Birrer} et~al.,}{{Birrer}
  et~al.}{2020}]{Birrer2020}
{Birrer} S.,  et~al., 2020, arXiv e-prints, \href
  {https://ui.adsabs.harvard.edu/abs/2020arXiv200702941B} {p. arXiv:2007.02941}

\bibitem[\protect\citeauthoryear{{Blandford} \& {Narayan}}{{Blandford} \&
  {Narayan}}{1992}]{Blandford1992}
{Blandford} R.~D.,  {Narayan} R.,  1992, \mn@doi [\araa]
  {10.1146/annurev.astro.30.1.311}, \href
  {https://ui.adsabs.harvard.edu/abs/1992ARA&A..30..311B} {30, 311}

\bibitem[\protect\citeauthoryear{{Chen} et~al.,}{{Chen}
  et~al.}{2016}]{Chen2016}
{Chen} G. C.~F.,  et~al., 2016, \mn@doi [\mnras] {10.1093/mnras/stw991}, \href
  {https://ui.adsabs.harvard.edu/abs/2016MNRAS.462.3457C} {462, 3457}

\bibitem[\protect\citeauthoryear{{Chen} et~al.,}{{Chen}
  et~al.}{2019}]{Chen2019}
{Chen} G. C.~F.,  et~al., 2019, \mn@doi [\mnras] {10.1093/mnras/stz2547}, \href
  {https://ui.adsabs.harvard.edu/abs/2019MNRAS.490.1743C} {490, 1743}

\bibitem[\protect\citeauthoryear{{Ciotti} \& {Bertin}}{{Ciotti} \&
  {Bertin}}{1999}]{C+B99}
{Ciotti} L.,  {Bertin} G.,  1999, \aap, \href
  {http://adsabs.harvard.edu/abs/1999A%26A...352..447C} {352, 447}

\bibitem[\protect\citeauthoryear{{Collett} \& {Cunnington}}{{Collett} \&
  {Cunnington}}{2016}]{Collett2016}
{Collett} T.~E.,  {Cunnington} S.~D.,  2016, \mn@doi [\mnras]
  {10.1093/mnras/stw1856}, \href
  {https://ui.adsabs.harvard.edu/abs/2016MNRAS.462.3255C} {462, 3255}

\bibitem[\protect\citeauthoryear{{Courbin} et~al.,}{{Courbin}
  et~al.}{2017}]{Courbin2017}
{Courbin} F.,  et~al., 2017, arXiv e-prints, \href
  {https://ui.adsabs.harvard.edu/abs/2017arXiv170609424C} {p. arXiv:1706.09424}

\bibitem[\protect\citeauthoryear{{Dai}, {Zackay}, {Venumadhav}, {Roulet}  \&
  {Zaldarriaga}}{{Dai} et~al.}{2020}]{Dai2020}
{Dai} L.,  {Zackay} B.,  {Venumadhav} T.,  {Roulet} J.,   {Zaldarriaga} M.,
  2020, arXiv e-prints, \href
  {https://ui.adsabs.harvard.edu/abs/2020arXiv200712709D} {p. arXiv:2007.12709}

\bibitem[\protect\citeauthoryear{{Dhawan} et~al.,}{{Dhawan}
  et~al.}{2020}]{Dhawan2020}
{Dhawan} S.,  et~al., 2020, \mn@doi [\mnras] {10.1093/mnras/stz2965}, \href
  {https://ui.adsabs.harvard.edu/abs/2020MNRAS.491.2639D} {491, 2639}

\bibitem[\protect\citeauthoryear{{Ding}, {Biesiada}  \& {Zhu}}{{Ding}
  et~al.}{2015}]{Ding2015}
{Ding} X.,  {Biesiada} M.,   {Zhu} Z.-H.,  2015, \mn@doi [\jcap]
  {10.1088/1475-7516/2015/12/006}, \href
  {https://ui.adsabs.harvard.edu/abs/2015JCAP...12..006D} {2015, 006}

\bibitem[\protect\citeauthoryear{{Ding} et~al.,}{{Ding}
  et~al.}{2017a}]{Ding2017}
{Ding} X.,  et~al., 2017a, \mn@doi [\mnras] {10.1093/mnras/stw3078}, \href
  {https://ui.adsabs.harvard.edu/abs/2017MNRAS.465.4634D} {465, 4634}

\bibitem[\protect\citeauthoryear{{Ding} et~al.,}{{Ding}
  et~al.}{2017b}]{Ding2017b}
{Ding} X.,  et~al., 2017b, \mn@doi [\mnras] {10.1093/mnras/stx1972}, \href
  {https://ui.adsabs.harvard.edu/abs/2017MNRAS.472...90D} {472, 90}

\bibitem[\protect\citeauthoryear{{Ding} et~al.,}{{Ding}
  et~al.}{2018}]{Ding2018}
{Ding} X.,  et~al., 2018, arXiv e-prints, \href
  {https://ui.adsabs.harvard.edu/abs/2018arXiv180101506D} {p. arXiv:1801.01506}

\bibitem[\protect\citeauthoryear{{Ding} et~al.,}{{Ding}
  et~al.}{2020}]{Ding2020}
{Ding} X.,  et~al., 2020, arXiv e-prints, \href
  {https://ui.adsabs.harvard.edu/abs/2020arXiv200608619D} {p. arXiv:2006.08619}

\bibitem[\protect\citeauthoryear{{Falco}, {Gorenstein}  \& {Shapiro}}{{Falco}
  et~al.}{1985}]{Falco1985}
{Falco} E.~E.,  {Gorenstein} M.~V.,   {Shapiro} I.~I.,  1985, \mn@doi [\apjl]
  {10.1086/184422}, \href
  {https://ui.adsabs.harvard.edu/abs/1985ApJ...289L...1F} {289, L1}

\bibitem[\protect\citeauthoryear{{Foreman-Mackey}, {Hogg}, {Lang}  \&
  {Goodman}}{{Foreman-Mackey} et~al.}{2013}]{Foreman-Mackey2013}
{Foreman-Mackey} D.,  {Hogg} D.~W.,  {Lang} D.,   {Goodman} J.,  2013, \mn@doi
  [\pasp] {10.1086/670067}, \href
  {https://ui.adsabs.harvard.edu/abs/2013PASP..125..306F} {125, 306}

\bibitem[\protect\citeauthoryear{{Freedman} et~al.,}{{Freedman}
  et~al.}{2019}]{Freedman2019}
{Freedman} W.~L.,  et~al., 2019, \mn@doi [\apj] {10.3847/1538-4357/ab2f73},
  \href {https://ui.adsabs.harvard.edu/abs/2019ApJ...882...34F} {882, 34}

\bibitem[\protect\citeauthoryear{{Freedman} et~al.,}{{Freedman}
  et~al.}{2020}]{Freedman2020}
{Freedman} W.~L.,  et~al., 2020, \mn@doi [\apj] {10.3847/1538-4357/ab7339},
  \href {https://ui.adsabs.harvard.edu/abs/2020ApJ...891...57F} {891, 57}

\bibitem[\protect\citeauthoryear{{Gilman}, {Birrer}  \& {Treu}}{{Gilman}
  et~al.}{2020}]{Gilman2020}
{Gilman} D.,  {Birrer} S.,   {Treu} T.,  2020, \mn@doi [\aap]
  {10.1051/0004-6361/202038829}, \href
  {https://ui.adsabs.harvard.edu/abs/2020A&A...642A.194G} {642, A194}

\bibitem[\protect\citeauthoryear{{Goldstein} \& {Nugent}}{{Goldstein} \&
  {Nugent}}{2017}]{Goldstein2017}
{Goldstein} D.~A.,  {Nugent} P.~E.,  2017, \mn@doi [\apjl]
  {10.3847/2041-8213/834/1/L5}, \href
  {https://ui.adsabs.harvard.edu/abs/2017ApJ...834L...5G} {834, L5}

\bibitem[\protect\citeauthoryear{{Goldstein}, {Nugent}, {Kasen}  \&
  {Collett}}{{Goldstein} et~al.}{2018}]{Goldstein2018}
{Goldstein} D.~A.,  {Nugent} P.~E.,  {Kasen} D.~N.,   {Collett} T.~E.,  2018,
  \mn@doi [\apj] {10.3847/1538-4357/aaa975}, \href
  {https://ui.adsabs.harvard.edu/abs/2018ApJ...855...22G} {855, 22}

\bibitem[\protect\citeauthoryear{{Grillo} et~al.,}{{Grillo}
  et~al.}{2018}]{Grillo2018}
{Grillo} C.,  et~al., 2018, \mn@doi [\apj] {10.3847/1538-4357/aac2c9}, \href
  {https://ui.adsabs.harvard.edu/abs/2018ApJ...860...94G} {860, 94}

\bibitem[\protect\citeauthoryear{{Grillo}, {Rosati}, {Suyu}, {Caminha},
  {Mercurio}  \& {Halkola}}{{Grillo} et~al.}{2020}]{Grillo2020}
{Grillo} C.,  {Rosati} P.,  {Suyu} S.~H.,  {Caminha} G.~B.,  {Mercurio} A.,
  {Halkola} A.,  2020, \mn@doi [\apj] {10.3847/1538-4357/ab9a4c}, \href
  {https://ui.adsabs.harvard.edu/abs/2020ApJ...898...87G} {898, 87}

\bibitem[\protect\citeauthoryear{{Hannuksela}, {Haris}, {Ng}, {Kumar}, {Mehta},
  {Keitel}, {Li}  \& {Ajith}}{{Hannuksela} et~al.}{2019}]{Hannuksela2019}
{Hannuksela} O.~A.,  {Haris} K.,  {Ng} K.~K.~Y.,  {Kumar} S.,  {Mehta} A.~K.,
  {Keitel} D.,  {Li} T.~G.~F.,   {Ajith} P.,  2019, \mn@doi [\apjl]
  {10.3847/2041-8213/ab0c0f}, \href
  {https://ui.adsabs.harvard.edu/abs/2019ApJ...874L...2H} {874, L2}

\bibitem[\protect\citeauthoryear{{Haris}, {Mehta}, {Kumar}, {Venumadhav}  \&
  {Ajith}}{{Haris} et~al.}{2018}]{Haris2018}
{Haris} K.,  {Mehta} A.~K.,  {Kumar} S.,  {Venumadhav} T.,   {Ajith} P.,  2018,
  arXiv e-prints, \href {https://ui.adsabs.harvard.edu/abs/2018arXiv180707062H}
  {p. arXiv:1807.07062}

\bibitem[\protect\citeauthoryear{{Holz}}{{Holz}}{2001}]{Holz2001}
{Holz} D.~E.,  2001, \mn@doi [\apjl] {10.1086/322947}, \href
  {https://ui.adsabs.harvard.edu/abs/2001ApJ...556L..71H} {556, L71}

\bibitem[\protect\citeauthoryear{{Huber}, {Suyu}, {Noebauer}, {Chan}, {Kromer},
  {Sim}, {Sluse}  \& {Taubenberger}}{{Huber} et~al.}{2020}]{Huber2020}
{Huber} S.,  {Suyu} S.~H.,  {Noebauer} U.~M.,  {Chan} J.~H.~H.,  {Kromer} M.,
  {Sim} S.~A.,  {Sluse} D.,   {Taubenberger} S.,  2020, arXiv e-prints, \href
  {https://ui.adsabs.harvard.edu/abs/2020arXiv200810393H} {p. arXiv:2008.10393}

\bibitem[\protect\citeauthoryear{{Kelly} et~al.,}{{Kelly}
  et~al.}{2015}]{Kelly2015}
{Kelly} P.~L.,  et~al., 2015, \mn@doi [Science] {10.1126/science.aaa3350},
  \href {https://ui.adsabs.harvard.edu/abs/2015Sci...347.1123K} {347, 1123}

\bibitem[\protect\citeauthoryear{{Kennedy} \& {Eberhart}}{{Kennedy} \&
  {Eberhart}}{1995}]{Kennedy1995}
{Kennedy} J.,  {Eberhart} R.,  1995, in Proceedings of ICNN'95 - International
  Conference on Neural Networks. pp 1942--1948 vol.4,
  \mn@doi{10.1109/ICNN.1995.488968}

\bibitem[\protect\citeauthoryear{{Knox} \& {Millea}}{{Knox} \&
  {Millea}}{2020}]{Knox2020}
{Knox} L.,  {Millea} M.,  2020, \mn@doi [\prd] {10.1103/PhysRevD.101.043533},
  \href {https://ui.adsabs.harvard.edu/abs/2020PhRvD.101d3533K} {101, 043533}

\bibitem[\protect\citeauthoryear{{Kochanek}}{{Kochanek}}{2002}]{Kochanek2002}
{Kochanek} C.~S.,  2002, \mn@doi [\apj] {10.1086/342476}, \href
  {https://ui.adsabs.harvard.edu/abs/2002ApJ...578...25K} {578, 25}

\bibitem[\protect\citeauthoryear{{Koopmans}, {Treu}, {Fassnacht}, {Blandford}
  \& {Surpi}}{{Koopmans} et~al.}{2003}]{Koopmans2003}
{Koopmans} L.~V.~E.,  {Treu} T.,  {Fassnacht} C.~D.,  {Blandford} R.~D.,
  {Surpi} G.,  2003, \mn@doi [\apj] {10.1086/379226}, \href
  {https://ui.adsabs.harvard.edu/abs/2003ApJ...599...70K} {599, 70}

\bibitem[\protect\citeauthoryear{{Koopmans} et~al.,}{{Koopmans}
  et~al.}{2009}]{Koo++09}
{Koopmans} L.~V.~E.,  et~al., 2009, \mn@doi [\apjl]
  {10.1088/0004-637X/703/1/L51}, \href
  {http://adsabs.harvard.edu/abs/2009ApJ...703L..51K} {703, L51}

\bibitem[\protect\citeauthoryear{{Krist}, {Hook}  \& {Stoehr}}{{Krist}
  et~al.}{2011}]{Krist2011}
{Krist} J.~E.,  {Hook} R.~N.,   {Stoehr} F.,  2011, {20 years of Hubble Space
  Telescope optical modeling using Tiny Tim}.
p. 81270J, \mn@doi{10.1117/12.892762}

\bibitem[\protect\citeauthoryear{{Li}, {Gao}, {Ding}, {Wang}  \& {Zhang}}{{Li}
  et~al.}{2018a}]{Li2018}
{Li} Z.-X.,  {Gao} H.,  {Ding} X.-H.,  {Wang} G.-J.,   {Zhang} B.,  2018a,
  \mn@doi [Nature Communications] {10.1038/s41467-018-06303-0}, \href
  {https://ui.adsabs.harvard.edu/abs/2018NatCo...9.3833L} {9, 3833}

\bibitem[\protect\citeauthoryear{{Li}, {Mao}, {Zhao}  \& {Lu}}{{Li}
  et~al.}{2018b}]{LiSS2018}
{Li} S.-S.,  {Mao} S.,  {Zhao} Y.,   {Lu} Y.,  2018b, \mn@doi [\mnras]
  {10.1093/mnras/sty411}, \href
  {https://ui.adsabs.harvard.edu/abs/2018MNRAS.476.2220L} {476, 2220}

\bibitem[\protect\citeauthoryear{{Liao}}{{Liao}}{2020}]{Liao2020}
{Liao} K.,  2020, \mn@doi [\apjl] {10.3847/2041-8213/abadfd}, \href
  {https://ui.adsabs.harvard.edu/abs/2020ApJ...899L..33L} {899, L33}

\bibitem[\protect\citeauthoryear{{Liao}, {Fan}, {Ding}, {Biesiada}  \&
  {Zhu}}{{Liao} et~al.}{2017}]{Liao2017}
{Liao} K.,  {Fan} X.-L.,  {Ding} X.,  {Biesiada} M.,   {Zhu} Z.-H.,  2017,
  \mn@doi [Nature Communications] {10.1038/s41467-017-01152-9}, \href
  {https://ui.adsabs.harvard.edu/abs/2017NatCo...8.1148L} {8, 1148}

\bibitem[\protect\citeauthoryear{{Liao}, {Biesiada}  \& {Fan}}{{Liao}
  et~al.}{2019}]{Liao2019}
{Liao} K.,  {Biesiada} M.,   {Fan} X.-L.,  2019, \mn@doi [\apj]
  {10.3847/1538-4357/ab1087}, \href
  {https://ui.adsabs.harvard.edu/abs/2019ApJ...875..139L} {875, 139}

\bibitem[\protect\citeauthoryear{{Liesenborgs} \& {De Rijcke}}{{Liesenborgs} \&
  {De Rijcke}}{2012}]{Liesenborgs2012}
{Liesenborgs} J.,  {De Rijcke} S.,  2012, \mn@doi [\mnras]
  {10.1111/j.1365-2966.2012.21751.x}, \href
  {https://ui.adsabs.harvard.edu/abs/2012MNRAS.425.1772L} {425, 1772}

\bibitem[\protect\citeauthoryear{{Millon} et~al.,}{{Millon}
  et~al.}{2020}]{Millon2020}
{Millon} M.,  et~al., 2020, \mn@doi [\aap] {10.1051/0004-6361/201937351}, \href
  {https://ui.adsabs.harvard.edu/abs/2020A&A...639A.101M} {639, A101}

\bibitem[\protect\citeauthoryear{{M{\"o}rtsell}, {Johansson}, {Dhawan},
  {Goobar}, {Amanullah}  \& {Goldstein}}{{M{\"o}rtsell}
  et~al.}{2020}]{Mortsell2020}
{M{\"o}rtsell} E.,  {Johansson} J.,  {Dhawan} S.,  {Goobar} A.,  {Amanullah}
  R.,   {Goldstein} D.~A.,  2020, \mn@doi [\mnras] {10.1093/mnras/staa1600},
  \href {https://ui.adsabs.harvard.edu/abs/2020MNRAS.496.3270M} {496, 3270}

\bibitem[\protect\citeauthoryear{{Navarro}, {Frenk}  \& {White}}{{Navarro}
  et~al.}{1997}]{Navarro1997}
{Navarro} J.~F.,  {Frenk} C.~S.,   {White} S. D.~M.,  1997, \mn@doi [\apj]
  {10.1086/304888}, \href
  {https://ui.adsabs.harvard.edu/abs/1997ApJ...490..493N} {490, 493}

\bibitem[\protect\citeauthoryear{{Oguri}}{{Oguri}}{2019}]{Oguri2019}
{Oguri} M.,  2019, \mn@doi [Reports on Progress in Physics]
  {10.1088/1361-6633/ab4fc5}, \href
  {https://ui.adsabs.harvard.edu/abs/2019RPPh...82l6901O} {82, 126901}

\bibitem[\protect\citeauthoryear{{Oguri} \& {Marshall}}{{Oguri} \&
  {Marshall}}{2010}]{Oguri2010}
{Oguri} M.,  {Marshall} P.~J.,  2010, \mn@doi [\mnras]
  {10.1111/j.1365-2966.2010.16639.x}, \href
  {https://ui.adsabs.harvard.edu/abs/2010MNRAS.405.2579O} {405, 2579}

\bibitem[\protect\citeauthoryear{{Petrushevska} et~al.,}{{Petrushevska}
  et~al.}{2016}]{Petrushevska2016}
{Petrushevska} T.,  et~al., 2016, \mn@doi [\aap] {10.1051/0004-6361/201628925},
  \href {https://ui.adsabs.harvard.edu/abs/2016A&A...594A..54P} {594, A54}

\bibitem[\protect\citeauthoryear{{Petrushevska}, {Okamura}, {Kawamata},
  {Hangard}, {Mahler}  \& {Goobar}}{{Petrushevska}
  et~al.}{2018a}]{Petrushevska2018a}
{Petrushevska} T.,  {Okamura} T.,  {Kawamata} R.,  {Hangard} L.,  {Mahler} G.,
   {Goobar} A.,  2018a, \mn@doi [Astronomy Reports]
  {10.1134/S1063772918120272}, \href
  {https://ui.adsabs.harvard.edu/abs/2018ARep...62..917P} {62, 917}

\bibitem[\protect\citeauthoryear{{Petrushevska} et~al.,}{{Petrushevska}
  et~al.}{2018b}]{Petrushevska2018}
{Petrushevska} T.,  et~al., 2018b, \mn@doi [\aap]
  {10.1051/0004-6361/201731552}, \href
  {https://ui.adsabs.harvard.edu/abs/2018A&A...614A.103P} {614, A103}

\bibitem[\protect\citeauthoryear{{Pierel}, {Rodney}, {Vernardos}, {Oguri},
  {Kessler}  \& {Anguita}}{{Pierel} et~al.}{2020}]{Pierel2020}
{Pierel} J.~D.~R.,  {Rodney} S.,  {Vernardos} G.,  {Oguri} M.,  {Kessler} R.,
  {Anguita} T.,  2020, arXiv e-prints, \href
  {https://ui.adsabs.harvard.edu/abs/2020arXiv201012399P} {p. arXiv:2010.12399}

\bibitem[\protect\citeauthoryear{{Pi{\'o}rkowska}, {Biesiada}  \&
  {Zhu}}{{Pi{\'o}rkowska} et~al.}{2013}]{Piorkowska2013}
{Pi{\'o}rkowska} A.,  {Biesiada} M.,   {Zhu} Z.-H.,  2013, \mn@doi [\jcap]
  {10.1088/1475-7516/2013/10/022}, \href
  {https://ui.adsabs.harvard.edu/abs/2013JCAP...10..022P} {2013, 022}

\bibitem[\protect\citeauthoryear{{Planck Collaboration} et~al.,}{{Planck
  Collaboration} et~al.}{2020}]{Planck2020}
{Planck Collaboration} et~al., 2020, \mn@doi [\aap]
  {10.1051/0004-6361/201833910}, \href
  {https://ui.adsabs.harvard.edu/abs/2020A&A...641A...6P} {641, A6}

\bibitem[\protect\citeauthoryear{{Refsdal}}{{Refsdal}}{1964}]{Refsdal1964}
{Refsdal} S.,  1964, \mn@doi [\mnras] {10.1093/mnras/128.4.307}, \href
  {https://ui.adsabs.harvard.edu/abs/1964MNRAS.128..307R} {128, 307}

\bibitem[\protect\citeauthoryear{{Riess}, {Casertano}, {Yuan}, {Bowers},
  {Macri}, {Zinn}  \& {Scolnic}}{{Riess} et~al.}{2021}]{Riess2021}
{Riess} A.~G.,  {Casertano} S.,  {Yuan} W.,  {Bowers} J.~B.,  {Macri} L.,
  {Zinn} J.~C.,   {Scolnic} D.,  2021, \mn@doi [\apjl]
  {10.3847/2041-8213/abdbaf}, \href
  {https://ui.adsabs.harvard.edu/abs/2021ApJ...908L...6R} {908, L6}

\bibitem[\protect\citeauthoryear{{Rusu} et~al.,}{{Rusu}
  et~al.}{2017}]{Rusu2017}
{Rusu} C.~E.,  et~al., 2017, \mn@doi [\mnras] {10.1093/mnras/stx285}, \href
  {https://ui.adsabs.harvard.edu/abs/2017MNRAS.467.4220R} {467, 4220}

\bibitem[\protect\citeauthoryear{{Rydberg}, {Whalen}, {Maturi}, {Collett},
  {Carrasco}, {Magg}  \& {Klessen}}{{Rydberg} et~al.}{2020}]{Rydberg2020}
{Rydberg} C.-E.,  {Whalen} D.~J.,  {Maturi} M.,  {Collett} T.,  {Carrasco} M.,
  {Magg} M.,   {Klessen} R.~S.,  2020, \mn@doi [\mnras]
  {10.1093/mnras/stz3203}, \href
  {https://ui.adsabs.harvard.edu/abs/2020MNRAS.491.2447R} {491, 2447}

\bibitem[\protect\citeauthoryear{{Saha}}{{Saha}}{2000}]{Saha2000}
{Saha} P.,  2000, \mn@doi [\aj] {10.1086/301581}, \href
  {https://ui.adsabs.harvard.edu/abs/2000AJ....120.1654S} {120, 1654}

\bibitem[\protect\citeauthoryear{{Saha} \& {Williams}}{{Saha} \&
  {Williams}}{2006}]{Saha2006}
{Saha} P.,  {Williams} L. L.~R.,  2006, \mn@doi [\apj] {10.1086/508798}, \href
  {https://ui.adsabs.harvard.edu/abs/2006ApJ...653..936S} {653, 936}

\bibitem[\protect\citeauthoryear{{Schneider}}{{Schneider}}{2006}]{Sch06}
{Schneider} P.,  2006, in {Meylan} G.,  {Jetzer} P.,  {North} P.,  {Schneider}
  P.,  {Kochanek} C.~S.,   {Wambsganss} J.,  eds, Saas-Fee Advanced Course 33:
  Gravitational Lensing: Strong, Weak and Micro. pp 1--89

\bibitem[\protect\citeauthoryear{{Schneider} \& {Sluse}}{{Schneider} \&
  {Sluse}}{2014}]{Schneider2014}
{Schneider} P.,  {Sluse} D.,  2014, \mn@doi [\aap]
  {10.1051/0004-6361/201322106}, \href
  {https://ui.adsabs.harvard.edu/abs/2014A&A...564A.103S} {564, A103}

\bibitem[\protect\citeauthoryear{{Schneider}, {Ehlers}  \& {Falco}}{{Schneider}
  et~al.}{1992}]{Schneider1992}
{Schneider} P.,  {Ehlers} J.,   {Falco} E.~E.,  1992, {Gravitational Lenses},
  \mn@doi{10.1007/978-3-662-03758-4.
}

\bibitem[\protect\citeauthoryear{{Shajib}, {Treu}, {Birrer}  \&
  {Sonnenfeld}}{{Shajib} et~al.}{2020a}]{Shajib2020b}
{Shajib} A.~J.,  {Treu} T.,  {Birrer} S.,   {Sonnenfeld} A.,  2020a, arXiv
  e-prints, \href {https://ui.adsabs.harvard.edu/abs/2020arXiv200811724S} {p.
  arXiv:2008.11724}

\bibitem[\protect\citeauthoryear{{Shajib} et~al.,}{{Shajib}
  et~al.}{2020b}]{Shajib2020}
{Shajib} A.~J.,  et~al., 2020b, \mn@doi [\mnras] {10.1093/mnras/staa828}, \href
  {https://ui.adsabs.harvard.edu/abs/2020MNRAS.494.6072S} {494, 6072}

\bibitem[\protect\citeauthoryear{{Shapiro}}{{Shapiro}}{1964}]{Shapiro1964}
{Shapiro} I.~I.,  1964, \mn@doi [\prl] {10.1103/PhysRevLett.13.789}, \href
  {https://ui.adsabs.harvard.edu/abs/1964PhRvL..13..789S} {13, 789}

\bibitem[\protect\citeauthoryear{{Shu}, {Bolton}, {Mao}, {Kang}, {Li}  \&
  {Soraisam}}{{Shu} et~al.}{2018}]{Shu2018}
{Shu} Y.,  {Bolton} A.~S.,  {Mao} S.,  {Kang} X.,  {Li} G.,   {Soraisam} M.,
  2018, \mn@doi [\apj] {10.3847/1538-4357/aad5ea}, \href
  {https://ui.adsabs.harvard.edu/abs/2018ApJ...864...91S} {864, 91}

\bibitem[\protect\citeauthoryear{{Suyu} et~al.,}{{Suyu}
  et~al.}{2013}]{Suyu2013}
{Suyu} S.~H.,  et~al., 2013, \mn@doi [\apj] {10.1088/0004-637X/766/2/70}, \href
  {https://ui.adsabs.harvard.edu/abs/2013ApJ...766...70S} {766, 70}

\bibitem[\protect\citeauthoryear{{Suyu} et~al.,}{{Suyu}
  et~al.}{2017}]{Suyu2017}
{Suyu} S.~H.,  et~al., 2017, \mn@doi [\mnras] {10.1093/mnras/stx483}, \href
  {https://ui.adsabs.harvard.edu/abs/2017MNRAS.468.2590S} {468, 2590}

\bibitem[\protect\citeauthoryear{{Suyu} et~al.,}{{Suyu}
  et~al.}{2020}]{Suyu2020}
{Suyu} S.~H.,  et~al., 2020, \mn@doi [\aap] {10.1051/0004-6361/202037757},
  \href {https://ui.adsabs.harvard.edu/abs/2020A&A...644A.162S} {644, A162}

\bibitem[\protect\citeauthoryear{{Tie} \& {Kochanek}}{{Tie} \&
  {Kochanek}}{2018}]{Tie2018}
{Tie} S.~S.,  {Kochanek} C.~S.,  2018, \mn@doi [\mnras]
  {10.1093/mnras/stx2348}, \href
  {https://ui.adsabs.harvard.edu/abs/2018MNRAS.473...80T} {473, 80}

\bibitem[\protect\citeauthoryear{{Tihhonova} et~al.,}{{Tihhonova}
  et~al.}{2018}]{Tihhonova2018}
{Tihhonova} O.,  et~al., 2018, \mn@doi [\mnras] {10.1093/mnras/sty1040}, \href
  {https://ui.adsabs.harvard.edu/abs/2018MNRAS.477.5657T} {477, 5657}

\bibitem[\protect\citeauthoryear{{Treu}}{{Treu}}{2010}]{Tre10}
{Treu} T.,  2010, \mn@doi [\araa] {10.1146/annurev-astro-081309-130924}, \href
  {http://adsabs.harvard.edu/abs/2010ARA%26A..48...87T} {48, 87}

\bibitem[\protect\citeauthoryear{{Treu} \& {Koopmans}}{{Treu} \&
  {Koopmans}}{2002a}]{TreuKoopmans2002}
{Treu} T.,  {Koopmans} L.~V.~E.,  2002a, \mn@doi [\mnras]
  {10.1046/j.1365-8711.2002.06107.x}, \href
  {https://ui.adsabs.harvard.edu/abs/2002MNRAS.337L...6T} {337, L6}

\bibitem[\protect\citeauthoryear{{Treu} \& {Koopmans}}{{Treu} \&
  {Koopmans}}{2002b}]{T+K02a}
{Treu} T.,  {Koopmans} L.~V.~E.,  2002b, \mn@doi [\apj] {10.1086/341216}, \href
  {http://adsabs.harvard.edu/cgi-bin/nph-bib_query?bibcode=2002ApJ...575...87T&db_key=AST}
  {575, 87}

\bibitem[\protect\citeauthoryear{{Treu} \& {Koopmans}}{{Treu} \&
  {Koopmans}}{2004}]{T+K04}
{Treu} T.,  {Koopmans} L.~V.~E.,  2004, \mn@doi [\apj] {10.1086/422245}, \href
  {http://adsabs.harvard.edu/abs/2004ApJ...611..739T} {611, 739}

\bibitem[\protect\citeauthoryear{{Treu} \& {Marshall}}{{Treu} \&
  {Marshall}}{2016}]{Treu2016}
{Treu} T.,  {Marshall} P.~J.,  2016, \mn@doi [\aapr]
  {10.1007/s00159-016-0096-8}, \href
  {https://ui.adsabs.harvard.edu/abs/2016A&ARv..24...11T} {24, 11}

\bibitem[\protect\citeauthoryear{{Wagner}}{{Wagner}}{2018}]{Wagner2018}
{Wagner} J.,  2018, \mn@doi [\aap] {10.1051/0004-6361/201834218}, \href
  {https://ui.adsabs.harvard.edu/abs/2018A&A...620A..86W} {620, A86}

\bibitem[\protect\citeauthoryear{{Wagner}, {Liesenborgs}  \&
  {Eichler}}{{Wagner} et~al.}{2019}]{Wagner2019}
{Wagner} J.,  {Liesenborgs} J.,   {Eichler} D.,  2019, \mn@doi [\aap]
  {10.1051/0004-6361/201833530}, \href
  {https://ui.adsabs.harvard.edu/abs/2019A&A...621A..91W} {621, A91}

\bibitem[\protect\citeauthoryear{{Wertz}, {Orthen}  \& {Schneider}}{{Wertz}
  et~al.}{2018}]{Wertz2018}
{Wertz} O.,  {Orthen} B.,   {Schneider} P.,  2018, \mn@doi [\aap]
  {10.1051/0004-6361/201732240}, \href
  {https://ui.adsabs.harvard.edu/abs/2018A&A...617A.140W} {617, A140}

\bibitem[\protect\citeauthoryear{{Witt}, {Mao}  \& {Keeton}}{{Witt}
  et~al.}{2000}]{Witt2000}
{Witt} H.~J.,  {Mao} S.,   {Keeton} C.~R.,  2000, \mn@doi [\apj]
  {10.1086/317201}, \href
  {https://ui.adsabs.harvard.edu/abs/2000ApJ...544...98W} {544, 98}

\bibitem[\protect\citeauthoryear{{Wong} et~al.,}{{Wong}
  et~al.}{2017}]{Wong2017}
{Wong} K.~C.,  et~al., 2017, \mn@doi [\mnras] {10.1093/mnras/stw3077}, \href
  {https://ui.adsabs.harvard.edu/abs/2017MNRAS.465.4895W} {465, 4895}

\bibitem[\protect\citeauthoryear{{Wong} et~al.,}{{Wong}
  et~al.}{2020}]{Wong2020}
{Wong} K.~C.,  et~al., 2020, \mn@doi [\mnras] {10.1093/mnras/stz3094}, \href
  {https://ui.adsabs.harvard.edu/abs/2020MNRAS.498.1420W} {498, 1420}

\bibitem[\protect\citeauthoryear{{Wucknitz}}{{Wucknitz}}{2002}]{Wucknitz2002}
{Wucknitz} O.,  2002, \mn@doi [\mnras] {10.1046/j.1365-8711.2002.05426.x},
  \href {https://ui.adsabs.harvard.edu/abs/2002MNRAS.332..951W} {332, 951}

\bibitem[\protect\citeauthoryear{{Yang}, {Ding}, {Biesiada}, {Liao}  \&
  {Zhu}}{{Yang} et~al.}{2019}]{Yang2019}
{Yang} L.,  {Ding} X.,  {Biesiada} M.,  {Liao} K.,   {Zhu} Z.-H.,  2019,
  \mn@doi [\apj] {10.3847/1538-4357/ab095c}, \href
  {https://ui.adsabs.harvard.edu/abs/2019ApJ...874..139Y} {874, 139}

\bibitem[\protect\citeauthoryear{{Yuan}, {Riess}, {Macri}, {Casertano}  \&
  {Scolnic}}{{Yuan} et~al.}{2019}]{Yuan2019}
{Yuan} W.,  {Riess} A.~G.,  {Macri} L.~M.,  {Casertano} S.,   {Scolnic} D.~M.,
  2019, \mn@doi [\apj] {10.3847/1538-4357/ab4bc9}, \href
  {https://ui.adsabs.harvard.edu/abs/2019ApJ...886...61Y} {886, 61}

\makeatother
\end{thebibliography}
\input{main.bbl}

\label{lastpage}

\end{document}